\documentclass[12pt]{article}

\usepackage{amsfonts}
\usepackage{epsfig}
\usepackage{latexsym}
\usepackage{graphicx}
\def\bequ{\begin{equation}}
\def\eequ{\end{equation}}
\def\barr{\begin{array}}
\def\earr{\end{array}}
\def\half{{1\over 2}}
\def\ben{\begin{equation}}
\def\een{\end{equation}}
\def\bena{\begin{eqnarray}}
\def\eena{\end{eqnarray}}

\renewcommand{\theequation}{\arabic{section}.\arabic{equation}}

\def\spa#1{\phantom{\fbox{\rule[-#1cm]{0cm}{0cm}}}}

\def\b1{e^0}

\newcommand{\be}{\begin{equation}}
\newcommand{\ee}{\end{equation}}
\def\bea{\begin{eqnarray}}
\def\eea{\end{eqnarray}}
\def\Tr{\mbox{Tr}}

\def\ket#1{| #1 \rangle}

\def\del {\partial}
\def\nn{\nonumber}

\def\half {{1 \over 2}}

\def\be{\begin{equation}}
\def\ee{\end{equation}}
\def\bea{\begin{eqnarray}}
\def\eea{\end{eqnarray}}

\def\lesssim{\mathrel{\hbox{\rlap{\hbox{\lower4pt\hbox{$\sim$}}}\hbox{$<$}}}}
\def\gtrsim{\mathrel{\hbox{\rlap{\hbox{\lower4pt\hbox{$\sim$}}}\hbox{$>$}}}}

\begin{document}
\hfuzz=100pt
\title{{\Large \bf{Classical and Quantum Strings in compactified pp-waves and G\"odel type Universes}}}
\author{\\Daniel Brace$^{1}$\footnote{brace@physics.technion.ac.il},\ Carlos A R Herdeiro$^{2}$\footnote{crherdei@fc.up.pt}, \ Shinji Hirano$^{1}$\footnote{hirano@physics.technion.ac.il}, 
\spa{0.5}\\
{$^{1}${\it Department of Physics, Technion}},
\\ {{\it Israel Institute of Technology}},
\\ {{\it Haifa 32000, Israel}}\spa{0.5}\\
{$^{2}${\it Centro de F\'\i sica do Porto}},
\\ {{\it Faculdade de Ci\^encias da Universidade do Porto}},
\\ {{\it Rua do Campo Alegre, 687, 4169-007 Porto, Portugal}}}

\date{July, 2003}
\maketitle


\begin{abstract} 
We consider Neveu-Schwarz pp-waves with spacetime supersymmetry. 
Upon compactification of a spacelike direction, 
these backgrounds develop Closed Null Curves (CNCs) and Closed Timelike Curves (CTCs), 
and are U-dual to supersymmetric G\"odel type universes. 
We study classical and quantum strings in this background, 
with emphasis on the strings winding around the compact direction. 
We consider two types of strings: \textit{long strings} stabilized by 
NS flux and \textit{rotating strings} which are stabilized against collapse by angular momentum. 
Some of the latter strings wrap around CNCs and CTCs, and are thus a potential source of pathology. 
We analyze the partition function, and in particular discuss the effects of these string states.
Although our results are not conclusive, the partition function seems to be dramatically altered due to the presence of CNCs and CTCs.
We discuss some interpretations of our results, including a possible sign of unitary violation.
\end{abstract}

\section{Introduction}
It was shown by Gauntlett et al \cite{Gauntlett:2002nw} that low energy string theory admits supersymmetric solutions of the G\"odel type \cite{Godel:ga}. These homogeneous spaces have Closed Timelike Curves (CTCs) and Closed Null Curves (CNCs) which are homotopic to a point. It was then pointed out that these Closed Causal Curves (CCCs) are not present in certain dimensional upliftings and certain T-dual versions of the G\"odel spaces \cite{Herdeiro:2002ft}. The reason was understood in \cite{Boyda:2002ba}: the uplifted or T-dual spacetime is a standard pp-wave, of the type that has been recently studied as the Penrose limit of certain near horizon brane geometries \cite{Berenstein:2002jq}. Such realization raised the hope that string theory could shed light on one of the most important open problems in General Relativity, or more generally, in gravitational theories: are geometries with CCCs intrinsically inconsistent? Or is propagation of matter in geometries with CCCs intrinsically inconsistent? The reason for this optimistic expectation concerning string theory and this problem is two-folded. 

On the one hand there are the hints left by the incomplete attempts to tackle this problem at the classical and semi-classical level. Classically there is a great number of solutions which obey all important energy conditions and have CCCs which are homotopic to a point. Lorentzian Taub-NUT is a striking example, since it is a completely non-singular vacuum solution. Thus such solutions cannot be ruled out by requirements on the matter content. Can we at least rule out classically \textit{creating} such solutions? Tipler claimed to have answered affirmatively to this question \cite{Tipler:76,Tipler:77}. However, he assumes the weak energy condition which might be violated in quantum processes. This led to several suggestions of `time machines' sourced by quantum effects, most notoriously the Morris-Thorne wormholes \cite{Morris:88} and to the proposal by Hawking that the spacetime `chronology protection' relies on quantum rather than classical effects \cite{Hawking:92}. Several works tried to establish that studying Quantum Field Theory in spacetimes with CCCs leads to inconsistencies, most importantly the violation of unitarity verified through the breakdown of the optical theorem. Examples include \cite{Sousa:89, Goldwirth:92, Boulware:92}. All these attempts are semi-classical.

On the other hand string theory is quantum gravity and we can quantize strings in the pp-waves dual to the G\"odel universe. Thus, these supersymmetric G\"odel solutions provide an ideal lab to test possible inconsistencies of quantum gravity in the presence of CCCs. This is the topic of this paper. Recent discussions of CTC's in string theory can be found in \cite{Boyda:2002ba, Harmark:2003ud, Dyson:2003zn, Gimon:2003ms, Biswas:2003ku, Drukker:2003sc, Hikida:2003yd, Brecher:2003rv}.

An immediate observation  concerns the choice of coordinates for the pp-wave. One can choose coordinates for which the pp-wave takes the canonical form, and the only non-trivial coefficient is $g_{uu}$. Then string quantization is equivalent to that of a set of massive scalar fields \cite{Jofre:hd, Metsaev:2001bj, Metsaev:2002re, Russo:2002rq, Blau:2003rt}. However, since quantization is in general performed in the light cone gauge, it is difficult to impose in these coordinates the identification that creates the non-trivial causal structure and gives the spacetime U-dual to G\"odel. By changing coordinates, string quantization becomes a Landau type problem \cite{Russo:1994cv, Russo:1995aj} (see some other related work \cite{David:2002km}). Then, the necessary identifications can be easily implemented. The price to pay is that the definition of quantum operators becomes dependent on the light cone momentum and the Hilbert space is naturally divided in a set of subspaces related by spectral flow \cite{Kiritsis:2002kz, D'Appollonio:2003dr}, which introduces some subtleties in the interpretation of some states and in the calculation of the partition function.

It is tempting to interpret the results herein along the lines
of some of the aforementioned semi-classical computations and 
other results in string theory: we discuss a possible signal of
 ghost states and thus unitarity violation. If this interpretation is
 correct, this example parallels the case of some supersymmetric rotating
 black holes. 
The BMPV black hole \cite{Breckenridge:96} has a CFT description in terms of the D1-D5 system. 
On the spacetime side there is a bound between angular momentum and mass; 
when the former exceeds the latter the event horizon disappears and the spacetime 
has CTCs through every point. On the CFT side, the states that one could associate 
to this 'over-rotating' case are ghosts \cite{Herdeiro:2000ap}. 
Note the  contrast with the dynamical violations of unitarity
 mentioned above for QFT in curved spacetimes with CTCs. Violating the
 unitarity at the level of the spectrum is more fundamental. It means that
 these spacetimes with closed timelike curves, albeit BPS states of
 supergravity, should not be states of string theory.

This paper is organized as follows. In section 2, we briefly review the
basics of the compactified pp-wave and G\"odel type solutions in string theory.
We then summarize the classical string solutions and their correspondence to
quantum string states in section 3. The classical string solutions will be
discussed in detail in section 4, first by using the Nambu-Goto string and then
by the Polyakov string. Section 5 is devoted to quantum strings, where we
discuss  the correspondence between classical and quantum states in
detail, and then analyze the partition function carefully. 
In section 6 we discuss some of our results; in particular we elaborate on the possible imprint of CCCs in the partition function.
In Appendix A, we show a supertube probe computation and make a clear
U-dual map to our string computation. Some details of the calculation in section 4 are shown in Appendix B. 
The reader can find the details of the
quantization of strings in this background in Appendix C. We briefly
summarize the Heisenberg algebra, $H_4$, and its associated spectral flow in Appendix D.

\section{Preliminary}
\setcounter{equation}{0}

\subsection{The pp-wave}
The solution of type II supergravity which will be studied was found by Horowitz and Tseytlin 
\cite{Horowitz:1994rf}. In Brinkmann form it reads\footnote{In this paper, we will assume $f$ to be positive.}
\bequ
\barr{c}
ds^2=dudv+2f(x^1dx^2-x^2dx^1)du+\delta_{ij}dx^idx^j \ , \spa{0.3}\\
B=- f(x^1dx^2-x^2dx^1)\wedge du \ , \label{htsol} \earr 
\eequ
where $i,j=3,\cdots,8$. 
This is a supersymmetric solution of IIA or IIB supergravity, 
preserving (at least) half of the vacuum supersymmetries, with Killing spinors which obey
\bequ
\Gamma^u\epsilon=0 \ . \eequ
Moreover, it is an exact solution of type II string theory to all orders in the string length. 
In fact, the only non-zero components of the curvature tensor are given by
\bequ
R_{u1u1}=R_{u2u2}=f^2 \ , \ \ \ \ R_{uu}=2f^2 \ , \ \ \ \ R=0 \ , \eequ
This implies that all scalar polynomials in the curvature vanish, 
since the curvature tensors are always proportional to a null vector. 
A similar argument holds for contractions of the  
curvature with the Neveu-Schwarz field. This background is 
also parallelizable, 
which means that the generalized curvature
\bequ
{\mathcal{R}}_{\mu \nu \alpha \beta}=R_{\mu \nu \alpha \beta}-\frac{1}{2}D_{\beta}H_{\alpha \mu \nu}+\frac{1}{2}D_{\alpha}H_{\beta \mu \nu}-\frac{1}{4}H_{\mu \beta \sigma}H_{\nu \alpha}^{\ \ \sigma}+\frac{1}{4}H_{\mu \alpha \sigma}H_{\nu \beta}^{\ \ \sigma} \ , \label{generalR} \eequ
vanishes, which implies the vanishing of the quartic fermionic terms in the worldsheet supersymmetric $\sigma$-model. Parallelizable pp-waves have been recently discussed in \cite{Sadri:2003ib}.

The spacetime (\ref{htsol}) is a pp-wave, since $\partial/\partial v$ is a covariantly constant null vector. 
Introducing polar coordinates $(r,\theta)$ in the $(x^1,x^2)$ plane and 
performing the coordinate transformation $\phi=\theta+fu$, we find (\ref{htsol}) takes the canonical pp-wave form.
\bequ
\barr{c}
ds^2=dudv-r^2f^2du^2+dr^2+r^2d\phi^2+\delta_{ij}dx^idx^j \ , \spa{0.3}\\
B=- fr^2d\phi\wedge du \ . \earr \label{ourstpp} \eequ
This geometry has also been studied by Nappi and Witten as an example of a 
WZW model \cite{Nappi:1993ie} based on a non-semisimple group.

\subsection{Closed Null Curves and Closed Timelike Curves}
The pp-wave gains an interesting causal structure when compactified in the direction of wave propagation. 
First, we define the spacelike and timelike coordinates 
\bequ
u=y-t \ , \ \ \ \ v=y+t \ , \eequ
and then take the $y$ direction to be compact with period $2\pi R$. 
The solution, using polar coordinates on the $(x^1,x^2)$ plane takes the form
\bequ
\barr{c}
ds^2=-dt^2+dy^2+2fr^2d\theta (dy-dt)+dr^2+r^2d\theta^2+\delta_{ij}dx^idx^j \ , \spa{0.3}\\
B=- fr^2d\theta\wedge (dy-dt) \ . \earr \label{ourst} \eequ 
The $\partial_y$ direction is always spacelike, and so is the $\partial_\theta$ direction. However, a linear combination of the two
\bequ
t^{\mu}\partial_ {\mu}=\alpha \partial_y+\frac{\beta}{r}\partial_{\theta} \ , \label{lc} \eequ has norm
\bequ
t^{\mu}t_{\mu}=\alpha^2\left(\frac{\beta}{\alpha}+fr+\sqrt{f^2r^2-1}\right)\left(\frac{\beta}{\alpha}+fr-\sqrt{f^2r^2-1}\right) \ . \label{norm} \eequ
Consider a curve 
at some fixed radius $r$ with tangent $t^{\mu}\partial_ {\mu}$. Such a curve winds around two periodic directions, $\theta$ and $y$. This curve can be closed as long as
\bequ
\frac{\beta}{\alpha}=\frac{r}{R}q \ , \ \ \ q \in {\mathbb{Q}} \ . \eequ
Thus, we find that this closed curve
has a tangent vector with norm 
\bequ
t^{\mu}t_{\mu}=\alpha^2f^2r^2\left(\frac{q}{Rf}+1+\sqrt{1-\frac{1}{r^2f^2}}\right)\left(\frac{q}{Rf}+1-\sqrt{1-\frac{1}{r^2f^2}}\right) \ . \eequ
\begin{description}
\item[$\bullet$] For $r<1/f$, this is a Closed \textit{Spacelike} Curve.
\item[$\bullet$] For $r=1/f$, this is a Closed \textit{Null} Curve if $fR \in {\mathbb{Q}}$.
\item[$\bullet$] For $r>1/f$, this is a Closed \textit{Timelike} Curve if 
\bequ
|q/Rf+1|< \sqrt{1-1/r^2f^2} \ . \eequ
\end{description}
We will refer to the surface defined by $r=1/f$ as the velocity of light surface (VLS). 
Notice that none of these closed curves are homotopic to a point and they disappear in the universal covering space of the manifold.



\subsection{Duality to a G\"odel type universe}
The solution (\ref{ourst}) is invariant under T-duality along the $y$ direction. However, by applying 
S-duality first (to get a IIB configuration) and then T-duality along the $y$-direction we arrive at the IIA configuration
\bequ
\barr{c}
\displaystyle{ds^2=-\left[dt+fr^2d\theta\right]^2+dy^2+dr^2+r^2d\theta^2+\delta_{ij}dx^idx^j} \ , \spa{0.5}\\
\displaystyle{B=fr^2dy\wedge d\theta \ , \ \ \ \ C^{(3)}=fr^2d\theta\wedge dt \wedge dy \ ,  \ \ \ \ C^{(1)}=-fr^2d\theta} \ . \earr \label{ourstIIA} \eequ
which  is a homogeneous spacetime. Since the vector 
$t^{\mu}\partial_{\mu}=\partial /\partial \theta$
becomes timelike when $r>1/f$, we also find CTCs in this spacetime. 
But now the CTCs are homotopic to a point, and remain closed in the covering space of the manifold.

\subsection{Kaluza-Klein reduction to a G\"odel type universe}
Again starting from the Horowitz-Tseytlin solution (\ref{ourst}) we now apply Kaluza-Klein reduction along the $y$ direction
and arrive at the nine dimensional configuration
\bequ
\barr{c}
\displaystyle{ds^2=-\left[dt+fr^2d\theta\right]^2+dr^2+r^2d\theta^2+\delta_{ij}dx^idx^j} \ , \spa{0.5}\\
\displaystyle{B=fr^2d\theta\wedge dt \ ,  \ \ \ \ A^{1}=A^{2}=-fr^2d\theta} \ . \earr \label{ourst9} \eequ
This is again a G\"odel type universe. The KK reduction produces the same effect as the dualities. That is, 
the CTCs in nine dimensions are homotopic to a point.

\section{Summary of Classical and Quantum Strings}
\setcounter{equation}{0}

Before going into details, we will summarize the results found in section 4.
We will consider five types of classical (un)stable string probes 
in the pp-wave description of the G\"odel-type spacetime (\ref{ourst}),
%
where the  longitudinal direction $y$ is compactified with radius $R$. We will consider strings which have winding $w$ around
the compact direction, momentum $m/R$ in the compact direction, and non-topological winding $w^{\prime}$ around the $\theta$
direction.

\subsection{Non-rotating Strings}

\bigskip\noindent
{\bf (i) Static Long Strings}: 

\medskip\noindent
These strings are stabilized due to the 
balance between the attractive force due to string tension and the 
repulsive force due to NS-NS flux. In fact, the potential in the radial direction is flat.
They can have arbitrary size at the same energy cost. The KK-momentum vanishes while the 
topological and non-topological windings, $w$ and $w'$, 
must satisfy $w'+2fRw=0$. 
As we will discuss further in Section 5, 
the zero modes associated to this flat potential lead to divergences in the partition function.
These strings are 
analogous to the long strings in $AdS_3$ found by 
Maldacena and Ooguri \cite{Maldacena:2000hw}, and in uncompactified 
Nappi-Witten background by Kiritsis and Pioline \cite{Kiritsis:2002kz}. 

\medskip\noindent
{\bf (ii) Expanding (or Contracting) Long Strings}: 

\medskip\noindent
This is a more general family of solutions containing the previous static strings as a special case. 
These strings do not feel any potential and can therefore expand or contract at constant velocity. 
The static strings are just the special case when the expansion velocity is zero. 

\medskip
As we will see below, these two types of strings only appear at special kinematical points characterized by
\begin{equation}
\vartheta\equiv \alpha'fp_+\ ,
\end{equation}
where $p_+$ is the \lq light-cone' momentum.
In particular, they appear when $\vartheta$ is an integer.

\begin{figure}
\begin{picture}(0,0)(0,0)
\put(-10,190){$H$}
\put(450,95){$r$}
\put(293,95){$r_{min}$}
\put(40,40){$r_{min}$}
\put(80,40){$r_{VLS}$}
\put(340,95){$r_{VLS}$}
\put(160,40){$r$}
\put(245,230){$H$}
\end{picture}   
\centering\epsfig{file=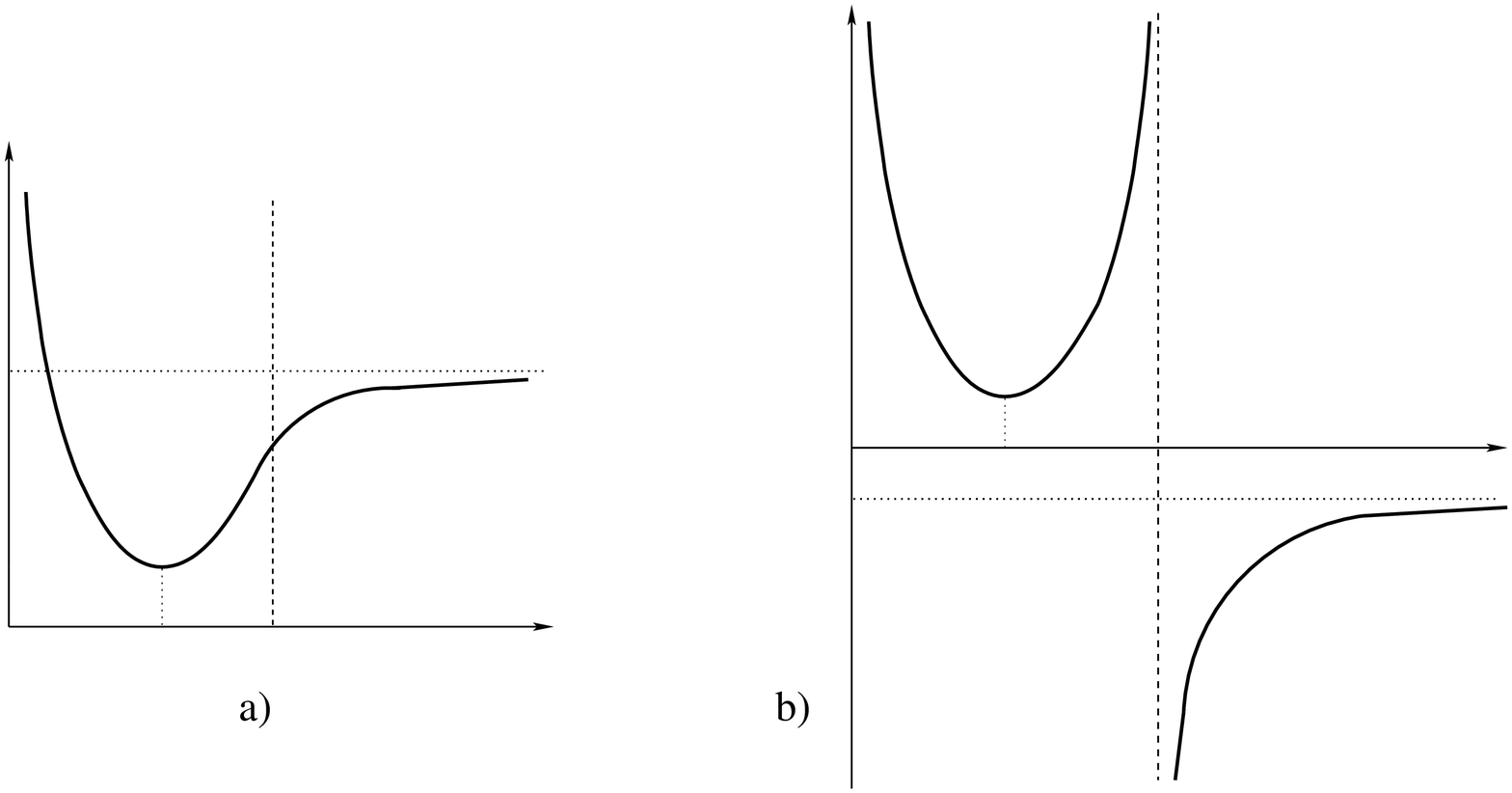,width=16cm}   
\caption{Effective potentials corresponding to either BPS or non-BPS strings.}
\label{ctccre}
\end{figure} 

\subsection{Rotating Strings}

We will distinguish 
three types of rotating strings,  two of which are stabilized by the balance of 
angular momentum and string tension. 
The third appears stable as a result of some kinematics that 
are peculiar to spacetimes with CNCs and CTCs.
There are two cases when we can study rotating strings with relative ease. 
They correspond to the parameter relations $m/R=Rw/\alpha'$ and $m/R=-Rww'/\alpha'(w'+2fRw)$.

\bigskip\noindent
{\bf (i) BPS String -- Supertube Dual}:

\medskip\noindent
These string configurations are U-dual to supertubes \cite{MT,Emparan:2001ux} 
on the spacetime (\ref{ourstIIA}). This identification is
explained further in Appendix A. 
These classical solutions have an energy and radius given by,
\bequ
E_{BPS} = \left| \frac{Rw}{\alpha'} \right| + \left| \frac{m}{R} \right| \ , \ \  
r^2_{BPS}= \left( \frac{\alpha'}{w'} \right)^2 \left| \left( \frac{Rw}{\alpha'}\right) \left( \frac{m}{R} \right) \right|, 
\label{stringBPS}
\eequ
which are independent of the flux $f$. 
The same solutions exist in flat space, and are known to be supersymmetric.
Although we have not checked the supersymmetry of these probes in this background,
we refer to these states as BPS since they have same energy and radius.
The potential felt by BPS strings can be any of 
Figure 1a), 1b) or 2a) and its radius any of the extrema in these figures. 
When the BPS string corresponds to the maximum of Figure 2a) 
it wraps around a CTC. Although this maximum might appear unstable, it actually is stable
\footnote{It is possible that $\sigma$ dependent radial pertubations might lead to instability, but a more thorough analysis is required to determine if this is so.}
for the following reason. 
In the G\"odel spacetime, beyond a certain radial point (necessarily after the VLS), 
the kinetic term of this string probe changes its sign, becoming negative. 
This coincides with the string passing from wrapping a standard CSC to wrapping a 
CNC and then a CTC. Heuristically, changing the sign of the kinetic term 
corresponds to changing the sign of the potential, since classical dynamics is 
not sensitive to the overall sign of the Lagrangian. This means that the string with
negative kinetic term wants to climb the potential. 
Considering the Hamiltonian dynamics leads to the same conclusion.
Arrows have been included in Figure 2 which indicate the natural sense
of motion due to the above argument.
The BPS strings that we will consider obey $m/R=Rw/\alpha'$.

\medskip
BPS strings only appear at the special kinematical point, $\vartheta=0$ (thus $p_+=0$), and correspond to right-moving excitations in the string spectrum.

\medskip\noindent
{\bf (ii) Non-BPS Strings}:

\medskip\noindent
The non-BPS strings are quite similar to BPS strings. 
but their energies and radii depend on the flux $f$. 
These strings can also correspond to any of the extrema in Figures 1a), 1b) and 2a). 
When they correspond to the maximum of Figure 2a), the non-BPS strings wrap around CTCs. 
The stabilization radii are $r_{non-BPS}=|Rw/(w'+2fRw)|$ for one case, and $r_{non-BPS}=|Rw/w'|$ for the other. 
The energies are given by $E=2|Rw(w'+2fRw)/(\alpha' w')|$ and $2Rw(w'+fRw)/(\alpha'(w'+2fRw))$, respectively.
When $f=0$, the radius and energy of non-BPS strings reduces to that of the BPS strings. 
Thus, turning on the flux $f$ breaks the degeneracy of the flat space string solutions. 

\medskip
The non-BPS strings correspond to right-moving excitations in one case and to 
left-moving excitations in the other, both at $\vartheta\ne 0$.

\begin{figure}
\begin{picture}(0,0)(0,0)
\put(0,190){$H$}
\put(434,75){$r$}
\put(117,77){$r_{max}$}
\put(40,77){$r_{min}$}
\put(71,77){$r_{VLS}$}
\put(280,75){$r_{VLS}$}
\put(170,77){$r$}
\put(210,190){$H$}
\end{picture}   
\centering\epsfig{file=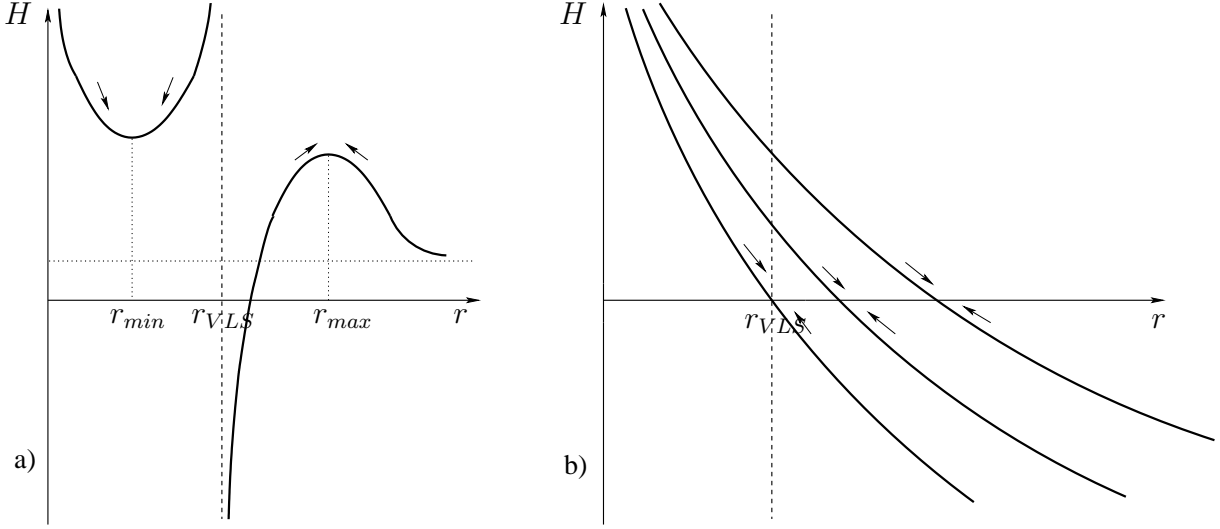,width=16cm}   
\caption{Effective potentials corresponding to BPS or non-BPS strings (left) and null strings (right).}
\end{figure}

\medskip\noindent
{\bf (iii) CNC Strings}:

\medskip\noindent
These are strings wrapping around CNCs. 
In contrast to the dynamical stability of being at the minimum (or \lq maximum') of the potential, 
the CNC strings are not at an extremum, but instead right in the middle of a slope. The potential
for several parameter values appears in Figure 2b). The point where the radial kinetic term changes sign corresponds
to the crossing of the horizontal line, which always occurs 
after the VLS.
Given the arguments concerning the stability of the maximum in Figure 2a) one might conclude that
the crossing points of Figure 2b) are also stable. In any case, these solutions show up in the
quantized string in Section 5, so are interesting to look at classically.
The CNC strings at \lq self-dual' compactification radius, $m/R=Rw/\alpha'$, 
have zero energy. When $m/R=-Rww'/\alpha'(w'+2fRw)$ the energy is given by $\pm 2f(Rw)^2/(\alpha'(w'+2fRw))$. In some cases the 
latter string 
actually obeys the BPS conditions (\ref{stringBPS}), but we will refer to it as a null string since its potential is quite
different from the other BPS strings.

\medskip
The zero energy CNC strings correspond to left-moving excitations in the spectral flowed representations of the quantized string, 
while the strings with non-zero energy correspond to right-moving excitations. The latter appears at $\vartheta=0$.

\section{Classical Strings}
\setcounter{equation}{0}

In this section we will clarify the statements outlined in the previous section by studying some simple string 
geodesics in the background (\ref{ourst}). 
We will work in the Nambu-Goto string first and then 
repeat the same analysis in the Polyakov string. 

\subsection{Nambu-Goto String}

The Nambu-Goto action is given by 
\begin{equation}
S=-{\cal T}\left[\int d\tau d\sigma\sqrt{-\det g_{\alpha\beta}}
+\int B_{ind}\right]\ ,
\label{NambuGoto}
\end{equation}
where ${\cal T}=1/(2\pi\alpha')$ is the string tension, and $g_{\alpha\beta}$ and $B_{ind}$ are the induced metric and two-form $B$-field respectively.

\medskip
\noindent
It is convenient to use the coordinates $y$ and $\psi=\theta+fy$, 
instead of $y$ and $\theta$.  Using static gauge, we consider solutions of the following form.
\begin{eqnarray}
t(\tau,\sigma)&=&\tau\ ,\\
y(\tau,\sigma)&=&2Rw\sigma+y(\tau)\ ,\\
\psi(\tau,\sigma)&=&2\Omega\sigma+\psi(\tau)\ ,\\
r(\tau,\sigma)&=&r(\tau)\ ,
\label{ansatz}
\end{eqnarray}
The remaining spacetime coordinates are taken to be constants. The range of $\sigma$ is from $0$ to $\pi$. 
At the end of the computation, we will replace $\Omega$ by $w'+fRw$, where 
$w'$ is the non-topological winding around the $\theta$ direction.
Using the background (\ref{ourst}) and the above ansatz the Nambu-Goto action (\ref{NambuGoto}) becomes
\begin{eqnarray}
S&=&-{\cal T}\int d\tau d\sigma \left[
2\sqrt{\frac{\Delta_{CNC}}{\Delta_{VLS}}
-\Delta_{CNC}\dot{r}^2
-{r^2 \over 4}\Delta_{VLS}\dot{\widetilde{\lambda}}^2}
-2\frac{\Omega fr^2}{\Delta_{VLS}}
+fr^2\dot{\widetilde{\lambda}}
\right]\ ,
\label{probeaction}
\end{eqnarray}
where we have defined 
\begin{eqnarray}
\dot{\widetilde{\lambda}}&\equiv&\dot{\lambda}
+2\frac{fRw+(\Omega-fRw)f^2r^2}{\Delta_{VLS}} \ , \\
\dot{\lambda}&\equiv&2(\Omega\dot{y}-Rw\dot{\psi})\ , \\
\Delta_{CNC}&\equiv&(Rw)^2+(\Omega^2-(fRw)^2)r^2 \ , 
\label{deltaCNC}\\
\Delta_{VLS}&\equiv&1-f^2r^2 \ .
\end{eqnarray}

At a fixed time $t$, the string is extended along the curve generated by the vector $\del/\del\sigma$. The norm of this vector can be calculated as follows. 
\begin{equation}
\frac{\del}{\del\sigma}={\del y \over \del\sigma}{\del \over \del y}
+{\del\theta \over \del\sigma}{\del \over \del\theta}
=2Rw{\del \over \del y}+2(\Omega-fRw){\del \over \del\theta} \equiv t^{\mu}\partial_{\mu}\ ,
\label{affine}
\end{equation}
Then, the norm $|\del/\del\sigma|^2=g_{\mu\nu}t^{\mu}t^{\nu}$  is given by
\begin{equation}
\left|\frac{\del}{\del\sigma}\right|^2=4\Delta_{CNC}\ .
\label{stringnorm}
\end{equation}
The quantity $\Delta_{CNC}$ determines whether the string wraps a spacelike ($\Delta_{CNC}>0$), 
null ($\Delta_{CNC}=0$), or timelike ($\Delta_{CNC}<0$) curve. 
$\Delta_{VLS}=0$ defines the velocity of light surface (VLS).
One can show $\Delta_{CNC}$ is strictly positive within the VLS. Defining $r_{CNC}$ as the radius at  which $\Delta_{CNC}=0$,
 we find,
\begin{equation}
r_{CNC}^2-{1 \over f^2}
=\frac{-\Omega^2}{f^2\left(\Omega^2-(fRw)^2\right)}
\ge 0\ ,\label{CNClocus}
\end{equation}
That last inequality follows since $\Omega^2-(fRw)^2=w'(w'+2fRw)$ must be less than zero in order 
for $\Delta_{CNC}$ to vanish somewhere.

One can also see that when the string wraps a timelike curve, the kinetic term $\Delta_{CNC}\dot{r}^2$ flips its sign. 
This typically implies that the timelike string behaves classically as if the potential were upside down. 
The negative kinetic term might be signaling that the dynamics of the string beyond CNC is pathological. 
That was the perspective taken in \cite{Drukker:2003sc} following \cite{Emparan:2001ux}. We will examine this point more carefully, in the context of
some specific examples. 

Since the canonical momentum  $p_y=\del {\cal L}/\del\dot{y}$ and $p_{\psi}=\del {\cal L}/\del\dot{\psi}$ conjugate 
to $y$ and $\psi$ are conserved quantities, we will consider the dynamics for 
fixed $p_y$ and $p_{\psi}$. As $\dot{y}$ and $\dot{\psi}$ 
only appear in the combination $\dot{\lambda}=2(\Omega \dot{y}-Rw\dot{\psi})$, 
it is convenient to take $\lambda$ as a dynamical variable. 
Then it is straightforward to compute the Hamiltonian  
${\cal H}=p_r\dot{r}+p_{\lambda}\dot{\lambda}-{\cal L}$, which is given by
\begin{eqnarray}
{\cal H}&=&2s{\cal T}
\sqrt{\frac{\Delta_{CNC}}{\Delta_{VLS}}\left(
1+\frac{(p_{\lambda}/{\cal T}+fr^2)^2}{r^2\Delta_{VLS}}
+\frac{(p_r/2{\cal T})^2}{\Delta_{CNC}}\right)}\\
&&\hspace{1.5cm}-2{\cal T}\Omega\frac{fr^2}{\Delta_{VLS}}
-\frac{2p_{\lambda}(fRw+(\Omega-fRw)f^2r^2)}{\Delta_{VLS}}\ ,\nn
\label{thehamiltonian}
\end{eqnarray}
where we defined
\begin{equation}
s=\mbox{sign}\left(\frac{\Delta_{CNC}}
{\Delta_{VLS}}\right)\ .
\end{equation}

\subsubsection{Non-Rotating Strings}

First we will consider non-rotating strings. Setting $p_{\lambda}=0$, the Hamiltonian reduces to 
\begin{eqnarray}
{\cal H}&=&2s{\cal T}
\sqrt{\frac{\Delta_{CNC}}{\Delta_{VLS}^2}
+\frac{(p_r/2{\cal T})^2}{\Delta_{VLS}}}
-2{\cal T}\Omega\frac{fr^2}{\Delta_{VLS}}\ .
\label{nonRhamilton}
\end{eqnarray}
A simple class of string solutions can be found when 
%
\begin{equation}
p_r=\pm {2{\cal T} \over f}\sqrt{\Omega^2-(fRw)^2}
\qquad\mbox{or}\qquad
\dot{r}=\pm{\sqrt{\Omega^2-(fRw)^2} \over \Omega}\ .
\label{velocity}
\end{equation}
With this solution, the energy is given by $\Omega/(\alpha' f)$. 

\bigskip\noindent
{\bf (i) Static Long Strings}

\medskip\noindent
When the condition, $\Omega^2-(fRw)^2=0$, is satisfied 
the string is static. Since the Hamiltonian (\ref{nonRhamilton}) is a constant the effective potential is flat and 
and the strings can have an arbitrary size at the same cost of energy.



\bigskip\noindent
{\bf (ii) Expanding (or Contracting) Long Strings}

\medskip\noindent
When $\Omega^2-(fRw)^2$ is not vanishing, the string has a constant velocity.
Since $\Omega=w'+fRw$, one can see that when the topological winding $w$ vanishes, the string expands at the velocity of light.

\medskip\noindent
In the Polyakov string in the next section, we will find more general non-rotating strings.

\subsubsection{Rotating Strings}

We will look for stable solutions with vanishing radial momentum. 
The effective potential for the radial mode 
is defined to be the Hamiltonian with $p_r=0$, and these are the potentials which are plotted in Figures 1 and 2.  

There are two cases where the square-root in the Hamiltonian (\ref{thehamiltonian}) completes:
\begin{equation}
p_{\lambda}=\pm{\cal T}\frac{Rw}{\Omega\mp fRw}\ .
\end{equation}

\paragraph{$\bullet$ The Case of $p_{\lambda}={\cal T}Rw/(\Omega- fRw)$}: 

In this  case, the condition on $p_{\lambda}$ can be rewritten in terms of $w$ and $w'$, 
\begin{equation}
{m \over R}={Rw \over \alpha'}\ ,
\label{selfdual}
\end{equation}
where we used the fact that $p_{\lambda}=p_y/(2w')-p_{\theta}/(Rw)$, and the fact that $p_{\theta}$ can be set to zero by 
reparametrization invariance. 
The relation (\ref{selfdual}) will be referred to as the \lq self-dual' radius condition. 

The Hamiltonian reduces to
\begin{equation}
H_{\pm}=\pm2\pi{\cal T}\frac{\Delta_{CNC}}{(\Omega-fRw)r(1\pm fr)}
\qquad \mbox{for}\quad|\Omega-fRw|
=\pm(\Omega-fRw)\ .
\label{rothamilton1}
\end{equation}
We look for extrema by calculating
\begin{eqnarray}
{\del H_{\pm} \over \del r}
= \pm 2\pi{\cal T}\frac{\left((\Omega-fRw)r\mp Rw\right)
\left((\Omega+fRw)r\pm Rw\right)}
{(\Omega-fRw)r^2(1\pm fr)^2}\ .
\label{SDextrema}
\end{eqnarray}
%
The shape of the \lq potential' depends on the range of parameters $\Omega$ and $w$. The following analysis is summarized
in Table 1.

\def\slash{\setlength{\unitlength}{1cm}\begin{picture}(0.5,0.5)(0.54,0.52)
\put(0.5,0.5){\line(2,1){0.71}}\end{picture}}
\def\nonBPS{$\mbox{BPS}\hspace{-0.73cm}\slash$}

\def\slash{\setlength{\unitlength}{1cm}\begin{picture}(0.5,0.5)(0.54,0.52)
\put(0.5,0.5){\line(2,1){0.71}}\end{picture}}
\def\nonBPS{$\mbox{BPS}\hspace{-0.73cm}\slash$}
\begin{table}[hbtp]
 \begin{center}
  \begin{tabular}{|c|c|c|c|c|c|c|c|c|c|}
   \hline
   \multicolumn{10}{|c|}
 {$p_{\lambda}={\cal T}Rw/(\Omega -fRw)$} \\
   \hline
   \multicolumn{4}{|c|}
 {(I) $\Omega^2-(fRw)^2>0$} &
   \multicolumn{6}{c|}
 {(II) $\Omega^2-(fRw)^2<0$} \\
   \hline
   \multicolumn{2}{|c|}{(A) $w>0$} &
   \multicolumn{2}{c|}{(B) $w<0$} &
   \multicolumn{4}{c|}{(C) $w>0$} &
   \multicolumn{2}{c|}{(D) $w<0$} \\
   \hline
  $\Omega>0$ & $\Omega<0$ & $\Omega>0$ & $\Omega<0$ &
  \multicolumn{2}{c|}{$\Omega>0$} &
  \multicolumn{2}{c|}{$\Omega<0$} & $\Omega>0$ & $\Omega<0$ \\
   \hline
   fig.1a & fig.1b & fig.1a & fig.1b &
  \multicolumn{2}{c|}{fig.2a} &
  \multicolumn{2}{c|}{fig.2a} & fig.2b & fig.2b \\
   \hline
   BPS & BPS & \nonBPS & \nonBPS & \nonBPS & BPS &
   \nonBPS & BPS & \nonBPS & \nonBPS \\
   \hline
   CSC & CSC & CSC & CSC & CSC & CTC &
   CTC & CSC & CNC & CNC \\
   \hline
   A$+$ & A$-$ & B$+$ & B$-$ & C$_{min}+$ & C$_{max}+$ &
   C$_{max}-$ & C$_{min}-$ & D$+$ & D$-$ \\
   \hline
  \end{tabular}
 \end{center}
 \caption{The classification of solutions for rotating strings with
$p_{\lambda}={\cal T}Rw/(\Omega- fRw)$: 
The first four rows indicate the parameter range of the solution. The potential for
each solution is plotted in the figure indicated in the fifth row. In the following
two rows, the BPS nature of the solution and the type of curve the string wraps is given.
The final row is a label for future reference.}
\end{table}

\medskip\noindent
(I) \underline{$\Omega^2-(fRw)^2>0$; strings which cannot wrap CNCs or CTCs}

\medskip\noindent
{\bf (A)} Let us consider the case $w>0$:


When $\Omega>0$, the condition, $\Omega^2-(fRw)^2>0$, implies $\Omega>fRw$. The corresponding Hamiltonian is then $H_+$, which 
has a single minimum at
\begin{equation}
r_{min}=\frac{Rw}{\Omega- fRw}=\frac{Rw}{w'}\ ,
\end{equation}
with energy given by
\begin{equation}
H_{min+}=4\pi{\cal T}Rw={2Rw \over \alpha'}\ .
\end{equation}
The potential is the one given by Figure 1a)
This solution is a BPS string, which is U-dual to a supertube.

\medskip\

When $\Omega<0$, the condition, $\Omega^2-(fRw)^2>0$, implies $\Omega<-fRw$. Since in this case, $\Omega-fRw<0$, the corresponding Hamiltonian is $H_-$, which has a single minimum at
\begin{equation}
r_{min}=-\frac{Rw}{\Omega- fRw}
=-\frac{Rw}{w'}\ ,
\end{equation}
with energy given by
\begin{equation}
H_{min-}=4\pi{\cal T}Rw={2Rw \over \alpha'}\ .
\end{equation}
The potential is the one in Figure 1b) with $r_{min}=-Rw/w'$. The solution at the minimum is again a BPS string.

\medskip\noindent
{\bf(B)} Next we turn to the case $w<0$:

\medskip\noindent
When $\Omega>0$, the condition, $\Omega^2-(fRw)^2>0$, implies $\Omega>-fRw$. Since $\Omega-fRw>0$, 
the corresponding Hamiltonian is $H_+$, which has a single minimum at
\begin{equation}
r_{min}=-\frac{Rw}{\Omega+ fRw}
=-\frac{Rw}{w'+ 2fRw}\ ,
\label{rmin1}
\end{equation}
with energy given by 
\begin{equation}
H_{min+}=-4\pi{\cal T}Rw\frac{\Omega+fRw}
{\Omega-fRw}
=-\frac{2Rw(w'+2fRw)}{\alpha' w'}\ .
\end{equation}
The potential is shown in  Figure 1a) with $r_{min}$ given by (\ref{rmin1}). The stable string at the minimum is non-BPS.

\medskip\noindent
When $\Omega<0$, the condition, $\Omega^2-(fRw)^2>0$, implies $\Omega<fRw$. Since $\Omega-fRw<0$, the corresponding Hamiltonian is $H_-$, which has a single minimum at
\begin{equation}
r_{min}=\frac{Rw}{\Omega+ fRw}
=\frac{Rw}{w'+ 2fRw}\ ,
\label{rmin2}
\end{equation}
and the minimum energy is again
\begin{equation}
H_{min-}=-4\pi{\cal T}Rw\frac{\Omega+fRw}
{\Omega-fRw}
=-\frac{2Rw(w'+2fRw)}{\alpha' w'}\ .
\end{equation}
The potential is shown in Figure 1b) with $r_{min}$ given by (\ref{rmin2}). The minimum again corresponds to a non-BPS string.

\medskip
As we will see below, both BPS and non-BPS string correspond to a right-moving excitation in the string spectrum.

\medskip\noindent
(II) \underline{$\Omega^2-(fRw)^2<0$; Strings which can wind CNCs and CTCs}

\medskip\noindent
{\bf (C)} First we consider the case $w>0$:

\medskip\noindent
For $\Omega>0$, the condition, $\Omega^2-(fRw)^2<0$, implies 
$\Omega<fRw$, thus the corresponding Hamiltonian is $H_-$,
which has a minimum at
\begin{equation}
r_{min}=\frac{Rw}{\Omega+ fRw}
=\frac{Rw}{w'+ 2fRw}\ ,
\end{equation}
and a maximum at
\begin{equation}
r_{max}=-\frac{Rw}{\Omega- fRw}
=-\frac{Rw}{w'}\ .
\end{equation}
The potential is shown in figure 2a) with the above values for the two extrema.
It can be shown that $r_{min}<r_{CNC}<r_{max}$. 
The maximum does not imply an instability, 
since the kinetic energy near $r_{max}$ is negative and 
the system is approximately an upside down harmonic oscillator 
with negative kinetic term. 
Thus the string wants to climb up the hill and 
is therefore stable at $r_{max}$. 
In fact, the string at $r_{max}$ is BPS. 
In terms of $w$ and $w'$, the radius and energy are given by
$-Rw/w'$ and $4\pi{\cal T}Rw$, respectively. 
Since $r_{max}>r_{CNC}$, this BPS string is wrapping a CTC. The string at $r_{min}$ is a non-BPS string.

\medskip\noindent
For $\Omega<0$, the condition, $\Omega^2-(fRw)^2<0$, implies 
$\Omega>-fRw$. Since $\Omega-fRw<0$, the corresponding Hamiltonian is again $H_-$. 
This case is almost the same as the $\Omega>0$ case, except that $r_{min}$ and $r_{max}$ are interchanged. 
The BPS string is now inside the VLS and wraps a spacelike curve, while the non-BPS string wraps around CTC.

\medskip\noindent
{\bf (D)} Next we consider the case $w<0$:

\medskip\noindent
For $\Omega>0$, the condition, $\Omega^2-(fRw)^2<0$, implies 
$\Omega<-fRw$. Since $\Omega-fRw>0$, the corresponding Hamiltonian is $H_+$. 

$\del H_+/\del r$ is negative, thus $H_+$ is given by Figure 2b). There appears to be no stable point. However, 
$H_+$ has a zero when $\Delta_{CNC}=0$. 
Three possibilities are plotted in Figure 2b). 
Since the kinetic term changes its sign as it crosses the CNC locus, 
the string actually begins to climb up to the CNC if it is initially placed beyond the CNC, 
while it begins to fall towards the CNC if it is placed inside the CNC point. 
The string wrapping around the CNC has zero energy.

\medskip
The zero energy CNC string corresponds to a left-moving excitation in the string spectrum.

\medskip\noindent
For $\Omega<0$, the condition, $\Omega^2-(fRw)^2<0$, implies 
$\Omega>fRw$. The corresponding Hamiltonian is again $H_+$. This case is identical to the  $\Omega>0$ case.

\paragraph{$\bullet$ The case of $p_{\lambda}=-{\cal T}Rw/(\Omega+ fRw)$}: 
This condition can be rewritten in terms of $w$ and $w'$ as
\begin{equation}
{m \over R}=-\frac{Rww'}{\alpha'(w'+2fRw)}\ .
\end{equation}
The Hamiltonian reduces to
\begin{equation}
H_{\pm}=\pm2\pi{\cal T}
\frac{(Rw)^2(1\pm fr)^2+\Omega^2r^2}{(\Omega+fRw)r(1\pm fr)}
\qquad \mbox{for}\quad|\Omega+fRw|
=\pm(\Omega+fRw)\ ,
\label{rothamilton2}
\end{equation}
and the extremum are defined by the zeros of
\begin{eqnarray}
{\del H_{\pm} \over \del r}
= \pm 2\pi{\cal T}\frac{\left((\Omega-fRw)r\mp Rw\right)
\left((\Omega+fRw)r\pm Rw\right)}
{(\Omega+fRw)r^2(1\pm fr)^2}\ .
\label{SDextrema2}
\end{eqnarray}
The behavior of the potential in this case is very similar to the previous one. The result is summarized in Table \ref{table2} and the details of the computation are shown in Appendix B:

\begin{table}[hbtp]
 \begin{center}
  \begin{tabular}{|c|c|c|c|c|c|c|c|c|c|}
   \hline
   \multicolumn{10}{|c|}
 {$p_{\lambda}=-{\cal T}Rw/(\Omega+fRw)$} \\
   \hline
   \multicolumn{4}{|c|}
 {(I) $\Omega^2-(fRw)^2>0$} &
   \multicolumn{6}{c|}
 {(II) $\Omega^2-(fRw)^2<0$} \\
   \hline
   \multicolumn{2}{|c|}{(K) $w>0$} &
   \multicolumn{2}{c|}{(L) $w<0$} &
   \multicolumn{2}{c|}{(M) $w>0$} &
   \multicolumn{4}{c|}{(N) $w<0$} \\
   \hline
  $\Omega>0$ & $\Omega<0$ & $\Omega>0$ & $\Omega<0$ &
  $\Omega>0$ & $\Omega<0$ &
  \multicolumn{2}{c|}{$\Omega>0$} &
  \multicolumn{2}{c|}{$\Omega<0$} \\
   \hline
   fig.1a & fig.1b & fig.1a & fig.1b & fig.2b & fig.2b &
  \multicolumn{2}{c|}{fig.2a} &
  \multicolumn{2}{c|}{fig.2a} \\
   \hline
   \nonBPS & \nonBPS & \nonBPS & \nonBPS & BPS & BPS
   & \nonBPS & \nonBPS &\nonBPS & \nonBPS \\
   \hline
   CSC & CSC & CSC & CSC & CNC & CNC &
   CSC & CTC & CTC & CSC \\
   \hline
   K$+$ & K$-$ & L$+$ & L$-$ & M$+$ & M$-$ &
   N$_{min}+$ & N$_{max}+$ & N$_{max}-$ & N$_{min}-$ \\
   \hline
  \end{tabular}
 \end{center}
 \caption{The classification of solutions for rotating strings with
$p_{\lambda}=-{\cal T}Rw/(\Omega+ fRw)$: The structure of this table is identical with that of Table 1.}
\label{table2}
\end{table}

%

\subsection{Polyakov String}

Having discussed some of the classical solutions of the Nambu-Goto string,
we now wish to do the same for the Polyakov
string.  Although the Nambu-Goto action elucidates the dynamics of the strings, 
the Polyakov action is more useful for comparing the classical string solutions 
with their quantum counterparts, which we will do in the next section. Since the string is most
easily quantized in light-cone gauge, we will make the same gauge choice in this section.
This gauge choice is not the same as that used in the Nambu-Goto action of the previous section, so the comparison
of the two classical solutions might not be straightforward. Nevertheless, for the solutions we consider,
the two approaches yield identical results.


We begin with the non-linear sigma model Lagrangian
 \begin{eqnarray}
{\mathcal{L}}&=&\frac{1}{2\pi\alpha '}\left[\partial_+
u\partial_-v+\partial_+v\partial_-u
+2if\del_+u
\left(X\partial_-\bar{X}-\bar{X}\partial_-X\right) 
\right.\nn\\
&&\left.
+\left(\partial_+X\partial_-\bar{X}
+\partial_-X\partial_+\bar{X}\right)
+\left(\partial_+\zeta^a\partial_-\bar{\zeta}^b
+\partial_-\zeta^a\partial_+\bar{\zeta}^b\right)
\delta_ {ab}\right] \ , 
\end{eqnarray}
which describes strings on the G\"odel type background. 
Here, $X$ is identified as $re^{i\theta}$ in the Nambu-Goto string of the previous section. 
Light cone gauge fixes
\bequ
u=u_0+\alpha'p_-\sigma_-+\alpha'p_+\sigma^+
\eequ
The non-zero modes of $v$ are then determined by Virasoro constraints.
The zero mode of $v$ is defined by
\bequ
v=v_0+\alpha' q_-\sigma^-+\alpha'q_+\sigma^+ \ . 
\eequ
As shown in Appendix C, with this gauge choice, $X$ has the following mode expansion.
\begin{eqnarray}
X=i\sqrt{{\alpha' \over 2}}\sum_{n=-\infty}^{\infty}
\left({1 \over n-\vartheta}\widetilde{\alpha}_{n-\vartheta}
e^{-2in\sigma^+}+{1 \over n+\vartheta}\alpha_{n+\vartheta}
e^{-2i(n+\vartheta)\sigma^--2i\vartheta\sigma^+}\right)\ ,
\label{generalsolution}
\end{eqnarray}
where $\vartheta=\alpha'fp_+ $.
The crucial point in the Polyakov string is to impose the Virasoro constraints. 
The Virasoro zero modes are given by
\begin{eqnarray}
L_0&=&{\alpha' \over 4}p_-q_-+{1 \over 4\pi\alpha'}
\int_0^{\pi}d\sigma\left[\del_-Y_-\del_-\bar{Y}_-+i\alpha' fp_-
(Y_-\del_-\bar{Y}_--\bar{Y}_-\del_-Y_-)\right]\ ,\\
\widetilde{L}_0&=&{\alpha' \over 4}p_+q_++{1 \over 4\pi\alpha'}
\int_0^{\pi}d\sigma\left[\del_+Y_+\del_+\bar{Y}_+-i\alpha' fp_+
(Y_+\del_+\bar{Y}_+-\bar{Y}_+\del_+Y_+)\right]\ ,
\end{eqnarray}
where $Y=e^{2i\vartheta\sigma^+}X$. Later we will find it convenient to write the zero mode 
Virasoro constraints in terms of 
the energy $E$, the momentum and winding around the compact direction 
$m/R$, $w$, and the right moving angular momentum in $X$ plane. 
They are related to zero mode parameters of $u$ and $v$ by 
\begin{eqnarray}
p_{\pm}&=&\left(E+{m \over R}\pm{Rw \over \alpha'}\right)\ ,\\
q_{\pm}&=&\left(-E+{m \over R}\pm{Rw \over \alpha'}+4fJ\right)\ ,\\
J&=&-{i \over 4\pi\alpha'}\int_0^{\pi}d\sigma
(Y_-\del_-\bar{Y}_--\bar{Y}_-\del_-Y_-)\ .
\end{eqnarray}

\subsubsection{Non-Rotating String}

As we have seen in the Nambu-Goto string, non-rotating strings have a 
flat potential, in contrast to most states which are bound by the pp-wave background. 
The zero mode associated to this flat potential can be identified with the extra zero mode that
appears in the expansion of $X$ (\ref{generalsolution}) when  $\vartheta$ is an integer.
%
\begin{equation}
X=\left(x_0+2\alpha' p_0\tau\right)e^{-2il\sigma^+}+\cdots\ .
\end{equation}
%
%
If we identify the non-topological winding $w'$ with $-l$, then
the non-rotating string we found in the Nambu-Goto approach corresponds to
\begin{equation}
r=x_0+2\alpha' p_0\tau\ ,\qquad \theta=2iw'\sigma^+\ ,
\label{nonrotpol}
\end{equation}
where we take $x_0$ and $p_0$ real. 
The Virasoro constraints for this solution are
\begin{eqnarray}
0&=&L_0+\widetilde{L}_0={\alpha' \over 2}\left(-E^2+\left({m \over R}\right)^2
+\left({Rw \over \alpha'}\right)^2+p_0^2\right)
\quad(\mbox{mass-shell condition})\ ,\\
0&=&L_0-\widetilde{L}_0=mw\qquad(\mbox{level-matching})\ .
\end{eqnarray}
The level matching condition implies that $m=0$ which is in accord with $p_{\lambda}=0$, in the Namu-Goto string. 

In the case of the static long string, $p_0=0$ and the energy\footnote{Note the energy in the Polyakov framework should be identified with minus the energy from the Nambu-Goto formalism.} is given by $E_{\pm}=\pm Rw/\alpha'$. 
Taking the  $E_+$ solution, we find $\vartheta=2fRw$, which is exactly the condition $w'+2fRw=0$ of the Nambu-Goto
static string.
The $E_-$ solution, which has $\vartheta=0$, is the string with vanishing non-topological 
winding $w'$, which is the static long string found in the previous section. 

For the expanding (or contracting) long  string, 
we found $p_0=\pm{1 \over \alpha' f}\sqrt{w'(w'+2fRw)}$ in the Nambu-Goto approach. 
Plugging this into the mass shell condition gives $E=\pm{1 \over \alpha' f}(w'+fRw)$, in agreement 
with the Nambu-Goto string solution.

\subsubsection{Rotating String}

Since we are interested in strings with non-topological winding 
$w'$ and fixed radius $r$, we will focus on chiral strings. In order to compare these solutions with
those from the previous section we will include references to Tables 1 and 2. For example, $N_{min}+$
refers to the solution in Table 2, which exists at the minimum of the potential given in Figure 2a).

\paragraph{The Left-moving String}: We consider solutions of the form
\begin{equation}
X=re^{2iw'\sigma^+}\ .
\end{equation}
The Virasoro constraints are then given by,
\begin{eqnarray}
&&E^2=\left({m \over R}-{Rw \over \alpha'}\right)^2
\qquad(\mbox{mass-shell condition})\ ,\\
&&
{1 \over \alpha'}r^2w'(w'+\vartheta)=-mw
\qquad(\mbox{level-matching})\ .
\end{eqnarray}
%
where
\begin{equation}
\vartheta=\left\{
\begin{array}{ll}
4\alpha' f{m \over R} & \qquad E={m \over R}-{Rw \over \alpha'}\\
2fRw & \qquad E=-\left({m \over R}-{Rw \over \alpha'}\right)\ .
\end{array}
\right.
\end{equation}
%

First we will consider strings of vanishing energy, 
which occurs at the \lq self-dual' radius, defined by $m/R=Rw/\alpha'$. 
Since $mw>0$, the level matching condition can be satisfied only when $w'(w'+\vartheta)<0$.
Taking $\vartheta = 2fRw$, the level matching condition gives the radius of the classical solution
%
\begin{equation}
r^2=-{(Rw)^2 \over w'(w'+2fRw)}\ , \ \ \ (D+,D-)
\end{equation}
We can then identify this left moving string with the zero energy CNC string found in the Nambu-Goto approach.
The case when  $\vartheta =4\alpha' f{m \over R}$ is related to this one by T-duality.

%
%

On the other hand, when the condition $m/R=-Rww'/(\alpha'(w'+2fRw))$ is satisfied, the radius and the energy become
\begin{equation}
r=
\left\{
\begin{array}{ll}
\left|{Rw \over w'}\right|\quad
&\quad E=-\frac{2Rw(w'+fRw)}{\alpha'(w'+2fRw)}\ , \ \ \ \ \ \ (K+, K-, N_{min}+, N_{min}-)\\
\left|{Rw \over w'+2fRw}\right|\quad&
\quad E=\frac{2Rw(w'+fRw)}{\alpha'(w'+2fRw)}\ , \ \ \ \ \ \ \ \ (L+, L-, N_{max}+, N_{max}-)
\end{array}
\right.
\end{equation}
These are the non-BPS strings we found before. In particular, when $mw$ is positive 
these correspond to strings that can wrap a CTC.

As we will see in the next section, for left moving states, $mw$ being positive 
implies that the number operator $\widetilde{N}$ of the left-moving 
oscillators takes a negative eigenvalue, and the corresponding quantum 
states are spectral flowed ones. This is a peculiar property of  
string theory in this background. Naively, the negative number 
operator seems quite pathological. In particular, 
there appears to be an infinite degeneracy of
states at every level.
However, we will see that these 
problems are taken care of by the regularization employed in the computation of the partition function.

\paragraph{The Right-moving String}: We consider solutions of the form
\begin{equation}
X=re^{-2i\vartheta\sigma^+-2i(w'+\vartheta)\sigma^-}\ .
\end{equation}
In this case the Virasoro constraints read,
\begin{eqnarray}
&&p_+\left(-E+{m \over R}+{Rw \over \alpha'}+4f{mw \over w'}\right)=0
\qquad(\mbox{mass-shell condition})\ ,\label{massright}\\
&&
{1 \over \alpha'}r^2w'(w'+\vartheta)=mw
\qquad(\mbox{level-matching})\label{levelright}\ .
\end{eqnarray}

First consider the $p_+=0$ solution at the \lq self-dual' radius, $m/R=Rw/\alpha'$. 
The radius and energy of the string are given by
\begin{equation}
r=\pm {Rw \over w'}\ ,\qquad
E=-{2Rw \over \alpha'}\ \ \ \ \ \ \ \ (A+,A-,C_{max}+,C_{max}-).
\end{equation}
This corresponds to the BPS string found in Nambu-Goto approach. 
Solutions with $p_+\ne 0$ correspond to the non-BPS strings found before, whose radius and energy are given by 
%
\begin{equation}
r=\left|{Rw \over w'+2fRw}\right|\ ,\qquad
E={2Rw(w'+2fRw) \over \alpha'w'}\ \ \ \ \ \ (B+, B-, C_{min}+, C_{min}-).
\end{equation}
%

Both types of strings can wrap CTCs. However, unlike the left-moving strings, the wrapping of the 
CTC has nothing to do with the negative number operator. 
In particular, as we will see in the next section,
the corresponding quantum states of the $p_+=0$ solution are not spectral flowed ones.

When $m/R=-Rww'/(\alpha'(w'+2fRw))$, the radius and the energy of the $p_+=0$ solution is given by
\begin{equation}
r^2=-\frac{(Rw)^2}{w'(w'+2fRw)}\ ,
\qquad
E=-\frac{2f(Rw)^2}{\alpha'(w'+2fRw)}\ \ \ \ \ \ \ (M+,M-) ,
\end{equation}
which are exactly the CNC locus and the energy of the CNC string. In fact, 
it turns out that the other branch of the solution (\ref{massright}) degenerates to the $p_+=0$ case. 
With this condition on the KK-momentum, the right-moving string 
corresponds to the CNC string with non-zero energy found before. 

For all the cases of right moving strings we have just studied and identified with Nambu-Goto solutions, 
$mw>0$ was a necessary condition. It follows that the right moving number operator is also positive.


\subsubsection{The Particle Geodesics}

The geodesics of particles can also be derived from the Polyakov string. 
In the point particle limit, there is no $\sigma$ dependence. 
In particular the mode expansion of $X$ (\ref{generalsolution}) reduces to
its zero frequency part:
\begin{equation}
X=X^L_0-X^R_0e^{-4i\vartheta\tau}\ ,
\label{particlegeodesic}
\end{equation}
where $\vartheta=E+m/R$ since the winding $w$ vanishes. 
The Virasoro constraints for a massless particle amount to
\begin{eqnarray}
\left(E+{m \over R}\right)
\left(-E+{m \over R}+4f^2\left(E+{m \over R}\right)(X^R_0)^2\right)=0
\ ,\label{particlevirasoro}
\end{eqnarray}
which is independent of $X^L_0$. A solution with $E+m/R=0$ has $\vartheta=0$, 
and corresponds to a \lq static' particle located at $X=X^L_0-X^R_0$ and moving in the $y$-direction. 
For a particle that starts from the origin, $X^L_0=X^R_0$, the solution is given by
\begin{equation}
X={i \over f}\sqrt{{E-{m \over R} \over E+{m \over R}}}
e^{-2i\vartheta\tau}\sin(2\vartheta\tau)\ .\label{geodesic}
\end{equation}
For vanishing KK-momentum, 
the massless particle just reaches VLS before heading back to the origin. 
States with positive KK-momentum are confined to orbits within the VLS.
However, for negative KK-momentum, 
the massless particle has orbits which take it arbitrarily far away from the origin.

\section{Quantum Strings}
\setcounter{equation}{0}


Strings which wrap around CNCs or CTCs are potentially problematic. 
For example, one might guess that the CNC and CTC strings 
correspond to massless and tachyonic excitations, since naively 
the CNC and CTC strings seem analogous to states coming from null and timelike compactifications. 
However, as we have seen in the previous sections, the CTC string 
does not correspond to a tachyonic mode, rather its energy is real. 
Although there exists CNC strings with vanishing energy at the \lq self-dual' radius,
there also exists other CNC strings 
with positive energy. 
Since the kinetic term for the radial mode of the CNC and CTC strings 
is zero and negative respectively, another possibility is that there are zero and negative 
norm quantum states associated 
with these strings. 
However, we will find that the quantum string theory on this spacetime 
has neither null nor negative norm states.
In fact, by  constructing the Hilbert space using spectral flowed representations, 
would-be negative norm states are removed. 
However, this construction does not remove the quantum
states corresponding to CNC and CTC strings, which as we will see, exist with positive 
norm in the quantum theory.

We begin this section by defining the string vacuum in terms of spectral flowed representations.
States in the quantum spectrum are then associated to the classical solutions we studied in the
previous two sections. Once this correspondence is made, we can investigate the contributions
to the partition function coming from the different types of strings.

\subsection{The Spectrum}

In general, classical string solutions will be  corrected quantum mechanically. 
For example, zero point energy contributions will shift the mass shell condition. 
In the bosonic string on this background, the zero point energy is $\half\vartheta(1-\vartheta)$, 
while in type II string theories, the zero point energies are given by $-\half+{\vartheta \over 2}$ 
in the NS sector and $0$ in the R-sector. 
To associate classical configurations with quantum states, we will focus on
states in the RR-sector, where the zero point energy vanishes.

String theories in this background have been quantized by Russo and 
Tseytlin \cite{Russo:1994cv, Russo:1995aj} in the \lq light-cone' 
gauge and also, in the noncompact case, by Nappi and Witten \cite{Nappi:1993ie} 
covariantly (see also \cite{D'Appollonio:2003dr} for recent progress). 
The details of the light-cone quantization are reviewed in Appendix C.

Before discussing the spectrum, we will briefly review the 
construction of the Hilbert space. 
Since the only difference from the flat spacetime case 
appears in the nontrivial two-dimensional part, 
$X(\tau,\sigma)$ and $\bar{X}(\tau,\sigma)$, of the compactified pp-wave, 
we will focus our attention there. We will also only consider the bosonic part
of the spectrum. As shown in Appendix C, the quantization boils down to
\bequ
[\widetilde{\alpha}_{m-\vartheta},\widetilde{\alpha}^{\ast}_{n-\vartheta}]
=2(m-\vartheta)\delta_{m,n} \ , \qquad 
[\alpha_{m+\vartheta},\alpha^{\ast}_{n+\vartheta}]
=2(m+\vartheta)\delta_{m,n}\ .
\label{commutation}
\eequ 
When the twist parameter $\vartheta=\alpha'fp_+$ goes to zero, 
the quantization is the same as that of flat spacetime, and the left and right vacuua are defined by
\begin{eqnarray}
&&\widetilde{\alpha}_{m-\vartheta}\widetilde{\ket{0}}=0\ ,
\quad\mbox{for}\quad m>0\ ,
\qquad\widetilde{\alpha}^{\ast}_{m-\vartheta}\widetilde{\ket{0}}=0\ ,
\quad\mbox{for}\quad m<0\ ,\\
&&\alpha_{m+\vartheta}\ket{0}=0\ ,
\quad\mbox{for}\quad m>0\ ,
\qquad\alpha^{\ast}_{m+\vartheta}\ket{0}=0\ ,
\quad\mbox{for}\quad m<0\ .
\label{flatvac}
\end{eqnarray}
%
However, when the twist parameter $\vartheta$ becomes larger ( smaller ) than some integer, 
the role of annihilation and creation operators is interchanged for some modes, 
as is evident from (\ref{commutation}). For a generic value of $\vartheta$, the vacuum is defined by
\begin{eqnarray}
&&\widetilde{\alpha}_{m-\vartheta}\widetilde{\ket{\vartheta}}=0\ ,
\quad\mbox{for}\quad m>\vartheta\ ,
\qquad\widetilde{\alpha}^{\ast}_{m-\vartheta}\widetilde{\ket{\vartheta}}=0\ ,
\quad\mbox{for}\quad m<\vartheta\\
&&\alpha_{m+\vartheta}\ket{\vartheta}=0\ ,
\quad\mbox{for}\quad m>-\vartheta\ ,
\qquad\alpha^{\ast}_{m+\vartheta}\ket{\vartheta}=0\ ,
\quad\mbox{for}\quad m<-\vartheta\ .
\label{specvac}
\end{eqnarray}
Notice that extra zero modes appear when $\vartheta=\mathbb{Z}$. 
If the vacuum were not redefined in this way, the string states would have had 
negative norm. 

This discrete jump of the vacuum is equivalent to using spectral flowed representations
 (see, for example, section 3.1 and 11 of \cite{D'Appollonio:2003dr}). 
In fact, although the full-fledged $H_4$ current algebra of this 
background is not manifest in the light-cone gauge, 
there is a smaller subalgebra which is invariant under spectral flow. This is discussed further in Appendix D. 
The vacua (\ref{specvac}) can then be viewed as the spectral flowed image of the original vacua (\ref{flatvac}). 

After defining the vacuum, the next step in the quantization is to impose the Virasoro constraints. 
The mass-shell condition of the type II string theory is given by
\begin{equation}
\left(E-2fJ\right)^2={4 \over \alpha'}
\left(\widetilde{N}-a\right)
+\left(Q_++2fJ\right)^2+\vec{p}^2\ ,
\label{mscondition}
\end{equation}
and the level-matching condition is
\begin{equation}
N-\widetilde{N}=mw\ ,
\label{lmcondition}
\end{equation}
where the normal ordering constant $a$
is $\half$ in the NS sector and $0$ in the R sector. 
The left moving momentum is defined by
\bequ
Q_+ = m/R+Rw/\alpha'\ ,
\eequ
and $\vec{p}$ is the  
momentum in the transverse 6-dimensional space. The number operators $N$ and 
$\widetilde{N}$, and the right-moving angular momentum $J$ 
are defined in Appendix C.

From now on we will focus on states in the RR-sector.

\paragraph{The Left-moving String}: The classical string with non-topological winding $w'$ takes the form
\begin{equation}
X=re^{2iw'\sigma^+}\ ,
\end{equation}
where the radius $r$ is given by
\begin{equation}
{1 \over \alpha'}r^2w'(w'+\vartheta)=-mw\ .
\end{equation}
This suggests that the corresponding quantum state is excited by the oscillator  
$\widetilde{\alpha}_{-(w'+\vartheta)}$ or $\widetilde{\alpha}^{\ast}_{-(w'+\vartheta)}$. 
Since there is no right-moving excitation, we have $N=J=0$. Then, 
the mass-shell and level-matching conditions become
\begin{eqnarray}
E^2&=&\left({m \over R}-{Rw \over \alpha'}\right)^2+\vec{p}^2\ ,\\
\widetilde{N}&=&-mw\ .
\end{eqnarray}
One expects a coherent state of such oscillators to look like the corresponding classical states. 
%
%
In order that the left-moving classical string wrap a CNC 
the \lq self-dual' radius condition $m/R=Rw/\alpha'$ was required.
This is quite parallel to the quantum case, when the left moving string has zero 
energy. 
However, as we mentioned before, this implies that the number operator $\widetilde{N}$
takes a negative eigenvalue. 
For concreteness, let us assume $w'>0$. 
In this case, the classical CNC string exists if $w'+2fRw(=w'+\vartheta)$ is negative. 
In the quantum description, naively,  the oscillator $\widetilde{\alpha}_{-(w'+\vartheta)}$ would be a creation operator, 
but it is actually an annihilation operator on the spectral flowed representation since $w'+\vartheta$ is now negative.
\begin{equation}
\widetilde{\alpha}_{n-\vartheta}\ket{\widetilde{\vartheta}}_R=0
\quad\mbox{for}\quad n>\vartheta\ ,\qquad
\widetilde{\alpha}^{\ast}_{n-\vartheta}\ket{\widetilde{\vartheta}}_R=0
\quad\mbox{for}\quad n<\vartheta\ ,
\end{equation}
Since $\widetilde{N}$ still annihilates the spectral flowed vacuum\footnote{This is true if the definition of the fermionic vacuum is changed in the same way as the bosonic vacuum. It is not clear that this is the correct construction. In any case, when calculating the partition function this choice does not seem to matter.}
and $[\widetilde{N},\widetilde{\alpha}_{-(w'+\vartheta)}] = -w' \widetilde{\alpha}_{-(w'+\vartheta)}$ we see that  
$\widetilde{N}$ has a negative eigenvalue on the state
%
%
$\widetilde{\alpha}^{\ast}_{-(w'+\vartheta)}\ket{\widetilde{\vartheta}}_R$.
The point is that the classical zero energy CNC string should correspond, in the classical limit, to a coherent state built from 
spectral flowed representations. This allows the number operator to become negative. As a result of this correspondence, we will
refer to these quantum states as CNC strings.

A similar argument holds for non-BPS strings with $m/R=-Rww'/(\alpha'(w'+2fRw))$.

\paragraph{The Right-moving String}: Similarly the right-moving classical string takes the form
\begin{equation}
X=re^{-2i\vartheta\sigma^+-2i(w'+\vartheta)\sigma^-}\ ,
\end{equation}
where the radius is given by
\begin{equation}
{1 \over \alpha'}r^2w'(w'+\vartheta)=mw\ .
\end{equation}
This suggests that the corresponding quantum state is excited either by the oscillator  $\alpha_{w'+\vartheta}$ or $\alpha^{\ast}_{w'+\vartheta}$. Let us assume that $w'>0$, and consider the state $\ket{\widetilde{\vartheta}}_R\otimes \left(\alpha^{\ast}_{w'+\vartheta}\right)^k\ket{\vartheta}_R$. This state has the $J$-eigenvalue $k$. Thus the mass-shell and level-matching condition become
\begin{eqnarray}
(E-2fk)^2&=&(Q_++2fk)^2+\vec{p}^2\ ,\\
N&=&mw\ .
\end{eqnarray}
This leads to the following two solutions for $\vec{p}=0$,
\begin{equation}
p_+=0\ , \qquad\mbox{or}\qquad
-E+Q_++4fk=0\ ,
\end{equation}
which is parallel to the classical case (\ref{massright}), since $N$ has eigenvalue $w'k$ on the above state. 
In this case, one should also be able to construct coherent states which reduce to 
the classical configurations in the classical limit. 
For the classical configurations we studied, one at the \lq self-dual' radius and the other 
when $m/R=-Rww'/(\alpha'(w'+2fRw))$, 
the number operator $N$ is positive as one can see from the level-matching condition.

As we discussed in the classical case, the string with $p_+=0$ 
corresponds to either the BPS string U-dual to a supertube or the CNC string with non-zero energy. 
The zero mode $p_+=0$ in the light-cone gauge is believed to encode the 
information of the background itself. Although, the gauge choice used here is not quite
light cone gauge, since $p_+$ depends on the winding $w$, it is interesting to note that 
the BPS 
string which is dual to a supertube occurs in the $p_+=0$ sector. 
In the construction of \cite{Drukker:2003sc} 
a supertube domain wall is used to create a background which 
in some bounded region reproduces the G\"odel background. 

As far as the spectrum is concerned, 
there is no restriction on the frequency $w'$ and the winding number $w$. 
In particular, the parameters could obey $w'(w'+2fRw)<0$, 
which is the condition for the classical strings to wrap CNCs and CTCs. 
As we will see in the next section, the form of the partition function is changed
dramatically as a result of contributions coming from such states. 

\subsection{The Partition Function}

By constructing the Hilbert space using spectral flowed representations, we have seen that
would-be negative norm states are removed. However,  
${L}_0$ and $\tilde{L}_0$ are no longer bounded from below in this construction,
since the number operators can now take negative values.
Starting with some physical state, it is
always possible to act on it with creation operators in such a way that the level
is unchanged. Thus, we will find an infinite degeneracy of physical states at each level.
Consider also the CNC string with vanishing energy, which is an excited string, 
and appears at arbitrarily high oscillation level. The contribution to loops 
coming from the infinite degeneracy of such states 
is not suppressed by any mass term, as is the case in flat space. 

We will see this pathology appear 
in the calculation of the partition function, before regularization. 
The partition function was already computed by Russo and 
Tseytlin \cite{Russo:1994cv, Russo:1995aj}. However, in obtaining the 
modular invariant partition function, it is necessary to adopt a specific 
regularization which removes the divergences associated with the unbounded number operator. 
It appears that the regularization has the same effect as not 
performing the spectral flow while keeping the norm of the states positive. 
After regularization, the divergences associated with the zero energy CNC string and the non-BPS 
CTC string are eliminated. The regularization does not remove the contributions coming from
BPS strings, which wrap CTCs or CNCs with nonzero energy and  $p_+=0$. 

The partition function $V_{1-loop}$ of typeIIA/B strings is given by
\begin{eqnarray}
V_{1-loop}&=&-\frac{1}{2}V_{R\times R^6}
\int_{-\infty}^{+\infty}\frac{dEd^6p}{(2\pi)^7}
\int_{F_0}\frac{d^2\tau}{\tau_2}
\left[\Tr_{NS\otimes NS}\left(P_{GSO}^{--}
q^{L_0^{NS}}\bar{q}^{\widetilde{L}_0^{NS}}\right)\right.\\
&&\left.\hspace{-2.5cm}
-\Tr_{NS\otimes R}\left(P_{GSO}^{-\pm}
q^{L_0^{NS}}\bar{q}^{\widetilde{L}_0^{R}}\right)
-\Tr_{R\otimes NS}\left(P_{GSO}^{+-}
q^{L_0^{R}}\bar{q}^{\widetilde{L}_0^{NS}}\right)
+\Tr_{R\otimes R}\left(P_{GSO}^{+\pm}
q^{L_0^{R}}\bar{q}^{\widetilde{L}_0^{R}}\right)
\right] \ , \nn
\end{eqnarray}
%
where $q=\exp{(2\pi i \tau)}$, $\tau=\tau_1+i\tau_2$, $P_{GSO}^{\pm\pm}=\frac{1\pm (-1)^F}{2}\frac{1\pm (-1)^{\widetilde{F}}}{2}$ is the GSO projection, and the integration is carried out over the fundamental domain
\bequ
F_0=\{\tau|-1/2 <\tau_1<1/2 , |\tau|\ge 1 \} \ . \eequ
There is a subtle point in carrying out the calculation. 
At first sight, we are not allowed to perform 
the integration over $E$ before taking the trace over oscillator states,
since the representations we are tracing over depend on $\vartheta$ and therefore $E$. 
Nevertheless it will turn out that after the regularization the 
trace over states takes the same form for any value of $\vartheta$, 
which legalizes integration over $E$ first.
To illustrate, it is enough to look at the bosonic nontrivial 4-dimensional part of the spacetime: 
%
\begin{equation}
Z^B(\vartheta)\equiv 
\Tr_{\vartheta}
\left(q^{N_B+\vartheta J_B}\bar{q}^{\widetilde{N}_B+\vartheta J_B}
\right)\ .
\end{equation}
%
%
In the range $0\le m<\vartheta<m+1$ ( with a similar result for $\vartheta<0$ ), we find
\begin{eqnarray}
Z^B(\vartheta)&=&
\left|1+q^{\vartheta}+q^{2\vartheta}+\cdots\right|^2
\prod_{n=1}^{\infty}
\left(1+q^{n+\vartheta}\bar{q}^{\vartheta}
+(q^{n+\vartheta}\bar{q}^{\vartheta})^2+\cdots\right)\\
&&\hspace{-1.7cm}\times q^{-\half m(m+1)+m\vartheta}
\bar{q}^{m\vartheta}\prod_{n=1}^m
\left(1+q^{\vartheta-n}\bar{q}^{\vartheta}
+(q^{\vartheta-n}\bar{q}^{\vartheta})^2+\cdots\right)\nn\\
&&\hspace{-1.7cm}\times 
\prod_{n=m+1}^{\infty}
\left(1+q^{n-\vartheta}\bar{q}^{-\vartheta}
+(q^{n-\vartheta}\bar{q}^{-\vartheta})^2+\cdots\right)
\prod_{n=1}^{\infty}\left(1+\bar{q}^n+\bar{q}^{2n}+\cdots\right)\nn\\
&&\hspace{-1.7cm}\times 
\bar{q}^{-\half m(m+1)}\prod_{n=1}^m
\left(1+\bar{q}^{-n}
+\bar{q}^{-2n}+\cdots\right)
\prod_{n=m+1}^{\infty}
\left(1+\bar{q}^n+\bar{q}^{2n}+\cdots\right)\ .\nn
\end{eqnarray}
The first factor of the first line comes from the trace over the Landau level oscillators 
$\alpha_{\vartheta}$ and $\alpha^{\ast}_{\vartheta}$.
The following three infinite products come from the trace over the right-moving spectral flowed states,
while the last three come from the left-moving states.



A few comments are in order. 
First of all, the series $(1+\bar{q}^{-n}+\bar{q}^{-2n}+\cdots)$ is not convergent, and in general
neither is $(1+q^{n-\vartheta}\bar{q}^{-\vartheta}+(q^{n-\vartheta}\bar{q}^{-\vartheta})^2+\cdots)$. 
The terms which contain such factors as $q^{-n}$ and $\bar{q}^{-n}$ 
come from tracing over states where the number operator takes negative eigenvalues. 
For a fixed level $k$, there is an infinite degeneracy 
of states. In other words, the coefficient of the term $q^k$, for example, 
is divergent. Also one might wonder the terms such
as $q^{-n}$ indicates the appearance of excited tachyonic states. 
This does not seem to be the case, since the number operator having negative 
eigenvalue depends on kinematics, and in particular the energy. 

These problems are taken care of by the following regularization \cite{Russo:1994cv,D'Appollonio:2003dr, Maldacena:2000hw}, which 
is tantamount to regarding each series 
as formally convergent. Then one finds 
\begin{equation}
Z^B_{reg}(\vartheta)=\left|{1 \over 1-q^{\vartheta}}\right|^2
\prod_{n=1}^{\infty}{1 \over 1-q^{n+\vartheta}\bar{q}^{\vartheta}}
\prod_{n=1}^{\infty}{1 \over 1-q^{n-\vartheta}\bar{q}^{-\vartheta}}
\prod_{n=1}^{\infty}\left({1 \over 1-\bar{q}^n}\right)^2\ .
\end{equation}
Alternatively, the regularization is the same as \lq undoing' the spectral 
flow while keeping the norm of the states positive. 

Having specified the regularization prescription, we are now able to perform the 
integration over $E$ first and then take the trace over the states. After performing a  
Poisson resummation, we have
\begin{eqnarray}
V_{1-loop}&=&{\cal N}\int_{F_0}{d^2\tau \over \tau_2^6}
{R \over \sqrt{\alpha'}}\sum_{w,\widetilde{w}=-\infty}^{\infty}
\exp\left(-{\pi R^2 \over \alpha'\tau_2}
\left|\widetilde{w}-w\tau\right|^2\right)
{\cal Z}^B{\cal Z}^F\ .
\end{eqnarray}
where the bosonic part ${\cal Z}^B$ is given by
\begin{equation}
{\cal Z}^B=\left|\prod_{n=1}^{\infty}(1-q^n)\right|^{-16}
\frac{2\pi fR\left(\widetilde{w}-w\tau\right)}
{\sin\left(2\pi fR\left(\widetilde{w}-w\tau\right)\right)}
\prod_{n=1}^{\infty}\frac{(1-q^n)^2}
{(1-\rho q^n)(1-\rho^{-1}q^n)}\ ,
\end{equation}
and $\rho=\exp\left(-4\pi ifR\left(\widetilde{w}-w\tau\right)\right)$. 
The factor $2\pi fR\left(\widetilde{w}-w\tau\right)/\tau_2$ 
was added by hand, which does not naturally come about in the operator formalism, 
but was derived from the path integral computation \cite{Russo:1994cv}.
${\cal Z}^F$ is the fermionic part of the partition function which 
vanishes due to spacetime supersymmetry.
\paragraph{The divergence}:

As discussed by Russo and Tseytlin \cite{Russo:1994cv} 
there is a divergence in the partition function due to the
extra zero modes when $\vartheta\in\mathbb{Z}$. To see this, let us look at the
factor $1/(1-\rho q^n)(1-\rho^{-1}q^n)$ which can be rewritten as
\begin{equation}
\frac{1}{(1-e^{-4\pi ifR\widetilde{w}}\exp\left[2\pi i(n+2fRw)\tau\right])
(1-e^{4\pi ifR\widetilde{w}}\exp\left[2\pi i(n-2fRw)\tau\right])}\ .
\label{divergence}
\end{equation}
When $n\pm 2fRw=0$ and $2fR\widetilde{w}\in\mathbb{Z}$, the denominator
vanishes and the partition function is divergent. This  condition is
equivalent to $2fR$ being rational.
The first factor comes from the states built by the oscillator
$\alpha^{\ast}_{n+\vartheta}$, while the second comes from $\alpha_{-n+\vartheta}$. As
we have seen, the frequency $n$ corresponds to the non-topological winding
number $w'$ of the classical string; $w'>0$ for $\alpha^{\ast}_{n+\vartheta}$
($w'=n$) and $w'<0$ for $\alpha_{-n+\vartheta}$ ($-w'=n$). For the static long 
string (non-rotating string), we found that $\vartheta=2fRw$ and $w'+2fRw=0$. This is exactly the
condition $n \pm 2fRw=0$, which when satisfied leads to a divergence in the partition function,
and which coincides with
either the oscillator $\alpha^{\ast}_{n+\vartheta}$ or $\alpha_{-n+\vartheta}$
becoming a zero mode. Therefore the static long string is precisely the one
responsible for the divergence in the partition function. Physically, the
static long string, being effectively tensionless, can take arbitrary size at the
same energy cost and thus gives rise to a divergence due to the
integration over the infinite configuration-space volume.

Thus this divergence is not because of strings wrapping around CCCs, since the static long string wraps around CSCs.

\section{Discussions}
\setcounter{equation}{0}
In this section we discuss some of our results; in particular we elaborate on
the possible imprint of CCCs in the partition function.
We require two important facts: (1) the non-topological
winding $w'$ of the classical strings is associated to the frequency $n$ of the
string excitation. Similar results were found previously by Maldacena and Ooguri
\cite{Maldacena:2000hw} in $AdS_3$. (2) For strings to wrap around CNCs and
CTCs, the condition $w'(w'+2fRw)<0$ must be obeyed.

Again we focus on the factor $1/(1-\rho q^n)(1-\rho^{-1}q^n)$ of
(\ref{divergence}) in the partition function. There is a qualitative difference in
the contribution to the partition function between states with $n\pm 2fRw>0$ and states with $n\pm 2fRw<0$. 
At the
transition point $n\pm 2fRw=0$, we saw that there is a divergence in the
partition function due to the infinite configuration-space degeneracy of the static long string.
With the correspondence of the non-topological winding number $w'$ and the
frequency $n$, one can see that the condition $n\pm 2fRw<0$ is equivalent to
\begin{equation}
w'(w'+2fRw)<0\ .
\end{equation}
This is exactly when the classical strings are allowed to wrap CCCs.
Note that, for $n\pm 2fRw<0$, the magnitude of $\exp\left[2\pi i(n\pm
2fRw)\tau\right]$ is greater than $1$. This implies that the
factor $1/(1-\rho q^n)(1-\rho^{-1}q^n)$ could become negative.
In other words, the correct convergent series expansion of the factor
$1/(1-\rho^{\pm 1} q^n)$ is
\begin{eqnarray}
-\rho^{\mp 1}q^{-n}\left(1+\rho^{\mp 1}q^{-n}+\rho^{\mp 2}q^{-2n}+
\cdots\right)\ .
\label{ghostexpansion}
\end{eqnarray}
On the other hand, if $n\pm 2fRw$ were positive, the series expansion would
have been the standard one,
\begin{eqnarray}
1+\rho^{\pm 1} q^{n}+\rho^{\pm 2}q^{2n}+\cdots\ .
\end{eqnarray} 
We would therefore like to associate this qualitative change with the 
presence of CCCs.

What would be the physical interpretation of this transition? We do not have a
clear answer for that question, but here we discuss a couple of possibilities:

The sign in front of the series expansion may have a physical meaning.
It determines the statistics of the string excitation.
The spin is not changed, which is determined by whether the excitation is in
the NS or R-sector. Thus naively, it looks as if there were ghosts propagating
in the loop.

However, we discussed that there are neither null nor negative norm states in
the Hilbert space of this string theory. Moreover the light-cone gauge is ghost
free, even though the light-cone gauge adopted here is not quite the standard
one. The partition function, although vanishing, also contains a factor which is in general complex, so it
is also difficult to determine the importance of the sign in (\ref{ghostexpansion}).

Another possibility, is that the light-cone gauge which was used to quantize the string theory is not an
appropriate gauge choice given the causal structure of the spacetime.  In this case, the quantum theory
might be similar to those appearing in some time-dependent 
orbifolds \cite{Balasubramanian:2002ry}, where it was concluded that it is not
possible to choose a ghost-free gauge, in which computations can be carried out strictly in terms of
states with positive norm.

Given
that the G\"odel type universes have causal pathologies, 
it would be tempting to interpret the above observation as a signal of
the appearance of ghosts which is indicative of a unitarity violation.
But this interpretation would require much
more than the observations made here.

\section *{Acknowledgments}
We would like to thank L. Cornalba, N. Drukker, B. Fiol, J. Sim$\acute{\mbox{o}}$n for discussions. We are especially grateful to O. Bergman for useful discussions and to M. Costa who collaborated with us in the initial part of this project. S.H. would like to thank the warm hospitality of Centro de F\'\i sica do Porto, Faculdade de Ci\^encias da Universidade do Porto, where a part of this work was done. D.B. and S.H. were in part supported by Israel Science Foundation under grant 
No. 101/01-1. C.H. is supported by the 
grant SFRH/BPD/2001 and this project in part supported by POCTI/FNU/38004/2001--FEDER (FCT, Portugal)

\setcounter{section}{1}
\renewcommand{\theequation}{\Alph{section}.\arabic{equation}}
\section*{Appendix A: U-duality and Supertubes}
\setcounter{equation}{0}

The classical string configurations that we have 
studied can also be discussed in the U-dual framework where
many of the extended string configurations correspond to expanded D2-branes.  
Let us recall the U-dual metric and background fields (\ref{ourstIIA}).
\bequ
\barr{c}
\displaystyle{ds^2=-\left[dt+fr^2d\theta\right]^2+dy^2+dr^2+r^2d\theta^2+\delta_{ij}dx^idx^j} \ , \spa{0.5}\\
\displaystyle{B_{NS}=fr^2dy\wedge d\theta \ , \ \ \ \ C^{(3)}=fr^2d\theta\wedge dt \wedge dy \ ,  \ \ \ \ C^{(1)}=-fr^2d\theta} \ . \earr \label{ourstIIA2} \eequ

In flat space, it was shown \cite{MT} that cylindrical $D2$ branes can be supported against collapse by angular momentum
generated by electric and magnetic fields on their worldvolume. Furthermore, these supertubes are
 $\frac{1}{4}$ BPS. More recently, it was shown \cite{Drukker:2003sc} that in the G\"odel background (\ref{ourstIIA2}) supertubes
continue to exist and preserve the same supersymmetry as the background.  

If we consider a cylindrical $D2$-brane extended in $y$ and wrapping the $\theta$ direction $N$ times, the dynamics are 
described by a $U(N)$ gauge theory. Restricting attention to the $U(1)$ zero mode 
components of the field strength and radial mode, the
probe is described by a Born-Infeld Lagrangian with the usual couplings to the background $RR$ fields. In this case
we find   
\bequ
{\cal{L}}_{D2} = -|N| \sqrt{ (-{\dot{r}}^2 +\Delta_{VLS}^{-1})( {r^2}\Delta_{VLS} + \bar{B}^2 ) - \bar{E}^2 {r^2}\Delta_{VLS} } 
- Nfr^2  + Nfr^2E
\eequ 
where
\bequ
\Delta_{VLS} = 1 -f^2 r^2 \ , \ \ 
\bar{B} = B - fr^2  \ , \ \
\bar{E} = E - { f\bar{B}}\Delta_{VLS}^{-1}
\eequ
and $E$ = $F_{0y}$ and $B$ are the electric and magnetic fields on the brane. We choose to consider configurations
with $F_{0 \,\theta} = 0$.  
The Hamiltonian is given by
\bequ
{\cal{H}} = \frac{s \, |N|}{r\Delta_{VLS}} \sqrt{ {P_r}^2 r^2 \Delta_{VLS} + ( r^2 \Delta_{VLS}+\bar{\Pi}^2 )
(r^2 \Delta_{VLS} + \bar{B}^2) }
+\frac{Nf}{\Delta_{VLS}}(r^2\Delta_{VLS} + \bar{\Pi}\bar{B})
\eequ
where
\bequ
P_r = N^{-1}\frac{\partial{\cal{L}}}{\partial \dot{r}} \ , \ \    \Pi = N^{-1}\frac{\partial{\cal{L}}}{\partial E} \ , \ \ \bar{\Pi} = \Pi - fr^2 \ , \ \ s = \mbox{sign}(r^2 \Delta_{VLS} + \bar{B}^2)
\eequ
A $D2$-brane with nonzero field strength can be considered as
a bound state of $D2$-branes, $D0$-branes, and fundamental strings. 
The  conjugate momentum $N\Pi$ is just the 
number of strings that wrap $y$ and  $NB$ is the number of
of $D0$-branes per unit length in the $y$ direction.
For the supersymmetric configuration where $N,B$, and $\Pi$ are all positive the supertube has radius and energy given by 
\bequ
r_{BPS} = \sqrt{\Pi B} \ , \ \ {\cal{H}}_{BPS} = N\left(\Pi + B\right)\ .
\label{BPStube}
\eequ

To show the equivalence of this $D2$-brane probe with the string probe (\ref{NambuGoto}) it is convenient to 
set $2{\cal{T}} = 1$. Then one can check that the 
Lagrangian (\ref{probeaction}) and the Hamiltonian (\ref{thehamiltonian}) are identical
to those appearing in this Appendix under the following identification.
 
\bequ
N \rightarrow -w' \ , \ \
B \rightarrow -\frac{Rw}{w'} \ , \ \
E \rightarrow \frac{ \dot{ \lambda }}{2w'} \ , \ \
\Pi \rightarrow -{2}p_{\lambda} \ , \ \
P_r \rightarrow \frac{p_r}{w'}
\eequ
When making the comparison it is useful to note that $\Delta_{CNC} = N^2(r^2 \Delta_{VLS} + {\bar{B}}^2)$.
There are a few cases when the Hamiltonian ( with $P_r = 0$ ) has simple form.
\begin{eqnarray}
\Pi &=& B\ ,\\
\Pi &=& \frac{-B}{1-2fB}\ ,\\
\Pi &=& (2f)^{-1} \ , \ \  B = 0\ .
\end{eqnarray}
These cases  
correspond to those studied in the main body of this paper. 

\addtocounter{section}{1}
\setcounter{equation}{0}
\section*{Appendix B: More Nambu-Goto Results}

\medskip\noindent
(I) \underline{$\Omega^2-(fRw)^2>0$; strings do not wind CNCs or CTCs}

\medskip\noindent
{\bf (K)} Let us consider the case $w>0$:

\medskip\noindent
For $\Omega>0$, the condition, $\Omega^2-(fRw)^2>0$, implies $\Omega-fRw>0$. In this case $\Omega+fRw>0$,  and the corresponding Hamiltonian is $H_+$, which has a unique minimum at 
\begin{equation}
r_{min}=\frac{Rw}{\Omega-fRw}=\frac{Rw}{w'}\ ,
\end{equation}
with energy given by
\begin{equation}
H_{min+}=4\pi{\cal T}\frac{Rw\Omega}{\Omega+fRw}
=\frac{2Rw(w'+fRw)}{\alpha'(w'+2fRw)}>0\ .
\end{equation}
The potential is given by Figure 1a). Although the stabilization 
radius is the same as that of the BPS string in the previous case, 
it does not satisfy the BPS condition (\ref{stringBPS}). 
Thus the minimum corresponds to a non-BPS string.

\medskip\noindent
For $\Omega<0$, we have $\Omega-fRw<0$. The condition $\Omega^2-(fRw)^2>0$ implies $\Omega+fRw<0$, 
and the corresponding Hamiltonian is $H_-$, which has a unique minimum at
\begin{equation}
r_{min}=-\frac{Rw}{\Omega-fRw}=-\frac{Rw}{w'}\ ,
\end{equation}
with energy given by
\begin{equation}
H_{min-}=4\pi{\cal T}\frac{Rw\Omega}{\Omega+fRw}
=\frac{2Rw(w'+fRw)}{\alpha'(w'+2fRw)}>0\ .
\end{equation}
The potential is now given by Figure 1b). Again the stabilization radius is the same as that of the BPS string, but the energy does not saturate the BPS condition (\ref{stringBPS}). Thus the minimum corresponds to a non-BPS string.

\medskip\noindent
{\bf (L)} Next we consider the case $w<0$:

\medskip\noindent
For $\Omega>0$, we have $\Omega-fRw>0$. The condition $\Omega^2-(fRw)^2>0$ implies $\Omega+fRw>0$, and the corresponding Hamiltonian is $H_+$, which has a unique minimum at
\begin{equation}
r_{min}=-\frac{Rw}{\Omega+fRw}=-\frac{Rw}{w'+2fRw}\ ,
\end{equation}
with energy given by
\begin{equation}
H_{min+}=-4\pi{\cal T}\frac{Rw\Omega}{\Omega+fRw}
=-\frac{2Rw(w'+fRw)}{\alpha'(w'+2fRw)}>0\ .
\end{equation}
The potential is shown in Figure 1a) and the string at the minimum is non-BPS.

\medskip\noindent
For $\Omega<0$, we have $\Omega+fRw<0$, and the condition $\Omega^2-(fRw)^2>0$ implies $\Omega-fRw<0$. 
The corresponding Hamiltonian is $H_-$, which has a unique minimum at
\begin{equation}
r_{min}=\frac{Rw}{\Omega+fRw}=\frac{Rw}{w'+2fRw}\ ,
\end{equation}
with energy given by
\begin{equation}
H_{min-}=-4\pi{\cal T}\frac{Rw\Omega}{\Omega+fRw}
=-\frac{2Rw(w'+fRw)}{\alpha'(w'+2fRw)}>0\ .
\end{equation}
The potential is shown Figure 1b) and the string at the minimum is again non-BPS.

\medskip
These non-BPS strings correspond to left-moving excitations.

\medskip\noindent
(II) \underline{$\Omega^2-(fRw)^2<0$; Strings winding CNCs and CTCs}

\medskip\noindent
{\bf (M)} The case of $w>0$:

\medskip\noindent
For $\Omega>0$, we have $\Omega+fRw>0$, and the condition $\Omega^2-(fRw)^2<0$ implies $\Omega-fRw<0$. 
The corresponding Hamiltonian is $H_+$. 

The potential is now Figure 2b). In this case, there is no dynamically stable point, and $H_+$ has no zeros. 
At the $\Delta_{CNC}=0$ locus, $H_+$ is equal to $H_{CNC+}=4\pi{\cal T}f(Rw)^2/(\Omega+fRw)=2f(Rw)^2/(\alpha'(w'+2fRw))$. 
But by the argument given before the CNC may be a stable point. 
However the CNC string, in this case, has positive energy.

\medskip\noindent
For $\Omega<0$, we have $\Omega-fRw<0$. Since $\Omega+fRw>0$, the corresponding Hamiltonian is $H_+$. 
This case falls into the same pattern as the previous case. There is a CNC string whose energy is positive.

\medskip
The non-zero energy CNC string is actually BPS, as we can see from the BPS condition (\ref{stringBPS}), 
and corresponds to a right-moving excitation at $\vartheta=0$.

\medskip\noindent
{\bf (N)} The case of $w<0$:

\medskip\noindent
For $\Omega>0$, we have $\Omega-fRw>0$. Since $\Omega+fRw<0$, the corresponding Hamiltonian is $H_-$, which has a minimum at
\begin{equation}
r_{min}=-\frac{Rw}{\Omega-fRw}=-\frac{Rw}{w'}\ ,
\end{equation}
with energy at the minimum given by
\begin{equation}
H_{min-}=4\pi{\cal T}\frac{Rw\Omega}{\Omega+fRw}
=\frac{2Rw(w'+fRw)}{\alpha'(w'+2fRw)}>0\ ,
\end{equation}
and a maximum at
\begin{equation}
r_{max}=\frac{Rw}{\Omega+fRw}=\frac{Rw}{w'+2fRw}\ ,
\end{equation}
with energy at the maximum given by
\begin{equation}
H_{max-}=-4\pi{\cal T}\frac{Rw\Omega}{\Omega+fRw}
=-\frac{2Rw(w'+fRw)}{\alpha'(w'+2fRw)}<0\ .
\end{equation}
The potential is shown in Figure 2a). 
One can show that $r_{min}<r_{CNC}<r_{max}$.  
The energy at $r_{CNC}$ is given by $H_{CNC-}=4\pi{\cal T}f(Rw)^2/(\Omega+fRw)=2f(Rw)^2/(\alpha'(w'+2fRw))<0$. 
This CNC string appears to satisfy BPS condition (\ref{stringBPS}) but the overall sign of the energy is wrong. In any case,
this solution is unstable. 
Both strings at the minimum and maximum are non-BPS. The string at $r_{max}$ wraps around CTC.

\medskip\noindent
For $\Omega<0$, we have $\Omega+fRw<0$. The condition $\Omega^2-(fRw)^2<0$ implies $\Omega-fRw>0$. 
The corresponding Hamiltonian is again $H_-$. 
The potential, shown is figure 2a) is the same as the previous case, except that $r_{min}$ and $r_{max}$ are interchanged. 
Again both strings are non-BPS. The one at $r_{min}$ wraps around a CTC.

\medskip
These non-BPS strings correspond to a left-moving excitations, 
and the BPS CNC string corresponds to a right-moving excitation at $\vartheta=0$.

\addtocounter{section}{1}
\setcounter{equation}{0}
\section*{Appendix C: Quantization of Strings}

String quantization in pp-wave backgrounds has been considered by 
many authors (see \cite{Jofre:hd} for some early references). 
Russo and Tseytlin \cite{Russo:1994cv,Russo:1995aj} first 
considered string quantization in a homogeneous pp-wave with a 
compact direction, which is now known to be U-dual to a G\"odel Universe.  
In \cite{Harmark:2003ud} the string spectrum on the T-dual G\"odel 
Universe was discussed. Here we will recap the basic facts of the computation.
Since the Ramond-Ramond fields vanish, the RNS formalism can be used. The world sheet supersymmetric action is
\bequ
\mathcal{S}=\mathcal{S}_B+\mathcal{S}_F \ , \eequ
with the usual bosonic part
\bequ {\mathcal{S}}_B=-\frac{1}{4\pi\alpha '}\int
d^2\sigma \left[\sqrt{h}h^{\alpha \beta}\partial_
{\alpha}x^{\mu}\partial_ {\beta}x^{\nu}g_{\mu
\nu}+\epsilon^{\alpha \beta}\partial_ {\alpha}x^{\mu}\partial_
{\beta}x^{\nu}B_{\mu \nu}\right] \ , \label{sigma} \eequ
and fermionic piece,
\bequ {\mathcal{S}}_F=\frac{1}{4\pi\alpha '}\int
d^2\sigma \left[ig_{\mu \nu}\bar{\psi}^{\mu}\left(\gamma^{\alpha}D_ {\alpha}\psi\right)^{\nu}+\frac{1}{8}{\mathcal{R}}_{\mu \nu \sigma \tau}\bar{\psi}^{\mu}(1+\gamma_P)\psi^{\sigma}\bar{\psi}^{\nu}(1+\gamma_P)\psi^{\tau}\right] \ . \eequ
where boundary terms have been neglected.
The fermionic piece is obtained by introducing superfields on the worldsheet \cite{Braaten:85}. The worldsheet light-cone 
coordinates are defined by 
\bequ \left\{
\barr{l} \sigma^+=\tau+\sigma \spa{0.2}\\ \sigma^-=\tau-\sigma
\earr \right.   , \ \ \ \Rightarrow \ \ \ \left\{ \barr{l}
\partial_+=(\partial_{\tau}+\partial_{\sigma})/2 \spa{0.2}\\ \partial_{-}=(\partial_{\tau}-
\partial_{\sigma})/2 \earr \right.  , \eequ
and target space complex coordinates, $X$, $\{\zeta^a\}$, $\psi^X$, $\{\psi^{\zeta^a}\}$, $a=1,2,3$,  
are given by
\bequ  X=x^1+ix^2 \ ,  \ \ \
\left\{ \barr{l} \zeta^1=x^3+ix^4 \spa{0.2}\\
\zeta^2=x^5+ix^6 \spa{0.2}\\ \zeta^3=x^7+ix^8  \earr \right. \ , \ \ \ \psi^X=\psi^1+i\psi^2 \ ,  \ \ \
\left\{ \barr{l} \psi^{\zeta^1}=\psi^3+i\psi^4 \spa{0.2}\\
\psi^{\zeta^2}=\psi^5+i\psi^6 \spa{0.2}\\ \psi^{\zeta^3}=\psi^7+i\psi^8  \earr \right. \ . \eequ
The actions become
\bequ {\mathcal{S}}_B=\int d^2\sigma {\mathcal{L}}_B \ , \ \ \ {\mathcal{S}}_F=\int d^2\sigma \left({\mathcal{L}}_{Right}+{\mathcal{L}}_{Left}\right) \ , \eequ \begin{eqnarray}
{\mathcal{L}}_B&=&\frac{1}{2\pi\alpha '}\left[\partial_+
u\partial_-v+\partial_+v\partial_-u
+2if\del_+u
\left(X\partial_-\bar{X}-\bar{X}\partial_-X\right) 
\right.\nn\\
&&\left.
+\left(\partial_+X\partial_-\bar{X}
+\partial_-X\partial_+\bar{X}\right)
+\left(\partial_+\zeta^a\partial_-\bar{\zeta}^b
+\partial_-\zeta^a\partial_+\bar{\zeta}^b\right)
\delta_ {ab}\right] \ , 
\end{eqnarray}
\bequ
\barr{c} 
\displaystyle{{\mathcal{L}}_{Right}=\frac{1}{4\pi\alpha'}\left[i\psi^{u}\partial_+\psi^{v}+i\psi^{v}\partial_+\psi^u+i\psi^X\partial_+\psi^{\bar{X}}+i\psi^{\bar{X}}\partial_+\psi^X+
i\psi^{\zeta^a}\partial_+\psi^{\bar{\zeta^a}}+
i\psi^{\bar{\zeta^a}}\partial_+\psi^{\zeta^a}\right.} \spa{0.4}\\ ~~~~~~~~~~~~ \displaystyle{\left.+f(\bar{X}\psi^X-X\psi^{\bar{X}})\partial_+\psi^u+f\psi^u\partial_+(\bar{X}\psi^X-X\psi^{\bar{X}})+2f\partial_+ u\left(\psi^X\psi^{\bar{X}}-\psi^{\bar{X}}\psi^{X}\right)\right]} \ , \earr \eequ
\bequ
\barr{c} 
\displaystyle{{\mathcal{L}}_{Left}=\frac{1}{4\pi\alpha'}\left[i\tilde{\psi}^{u}\partial_-\tilde{\psi}^{v}+i\tilde{\psi}^{v}\partial_-\tilde{\psi}^u+i\tilde{\psi}^X\partial_-\tilde{\psi}^{\bar{X}}+i\tilde{\psi}^{\bar{X}}\partial_-\tilde{\psi}^X+i\tilde{\psi}^{\zeta^a}\partial_-\tilde{\psi}^{\bar{\zeta^a}}+i\tilde{\psi}^{\bar{\zeta^a}}\partial_-\tilde{\psi}^{\zeta^a}\right.} \spa{0.4}\\ ~~~~~~~~~~ \displaystyle{\left.+f(\bar{X}\tilde{\psi}^X-X\tilde{\psi}^{\bar{X}})\partial_-\tilde{\psi}^u+f\tilde{\psi}^u(\bar{X}\partial_-\tilde{\psi}^X-X\partial_-\tilde{\psi}^{\bar{X}})+3f(\tilde{\psi}^X\partial_-\bar{X}-\tilde{\psi}^{\bar{X}}\partial_-X)\tilde{\psi}^u\right]} \ .  \earr \eequ
The independent equations of motion are \bequ \barr{c}
\displaystyle{\partial_-\partial_+u=0} \ ,
\spa{0.4}\\
\displaystyle{\partial_+\partial_-X=-2if\partial_+u\partial_-X}
\ ,
\spa{0.3}\\
\displaystyle{\partial_-\partial_+v=-if\partial_+\left[X\partial_-\bar{X}-\bar{X}\partial_-X\right]}
\ , \spa{0.5}\\
\displaystyle{\partial_+\partial_-\zeta^a=0 }\ , \earr
\label{eqmot2} \eequ 

\bequ 
\begin{array}{ll}
i\partial_-\tilde{\psi}^v=f\left(\tilde{\psi}^X\partial_-\bar{X}-\tilde{\psi}^{\bar{X}}\partial_-X-\bar{X}\partial_-\tilde{\psi}^X+X\partial_-\tilde{\psi}^{\bar{X}}\right)\ ,&\quad
i\partial_+\psi^v=-f\partial_+(\bar{X}\psi^X-X\psi^{\bar{X}})\ ,\\
\partial_-\tilde{\psi}^u=0\ , &\quad
\partial_+\psi^u=0\ ,\\
i\partial_-\tilde{\psi}^X=2f\tilde{\psi}^u\partial_-X \ , &
\quad i\partial_+\psi^X=2f\psi^X\partial_+u\ ,\\
\partial_-\tilde{\psi}^{\zeta^a}=0 \ , &
\quad\partial_+\psi^{\zeta^a}=0\ , 
\end{array}
\label{eqmotfer} 
\eequ
Only when $f=0$ are these all free wave equations.
Choosing \lq light cone' gauge, 
\bequ
\barr{c}
u=u_0+\alpha'p_-\sigma_-+\alpha'p_+\sigma^+ \ , \spa{0.3}\\
\psi^u=0=\tilde{\psi}^u \ , \earr \eequ
the left-moving fermions become free. The non-zero modes 
$v$, $\psi^{v}$ and $\tilde{\psi}^v$ are then determined in terms of the transverse fields. 
The equations of motion for $X$ and $\psi^{X}$ are still non-trivial,
but by introducing new variables $Y$ and $\lambda$ 
\bequ
Y=e^{2i\vartheta\sigma^+}X \ , \ \ \ \ \ \lambda=e^{2i\vartheta\sigma^+}\psi^X \ , \label{freev} \eequ
where
\bequ
\vartheta=\alpha'fp_+ \ , \eequ
the wave equations become free.
The closed string periodicity condition 
for $X$ and $\psi^X$ becomes twisted in terms of the free variables.
\bequ
X(\tau, \sigma+\pi)=X(\tau, \sigma) \  \ \ \Leftrightarrow \ \ \ Y(\tau,\sigma+\pi)=e^{2\pi i \vartheta}Y(\tau,\sigma) \ , \eequ
\bequ
\psi^X(\tau, \sigma+\pi)=\pm \psi^X(\tau,\sigma) \ \ \ \Leftrightarrow \ \ \ \lambda(\tau, \sigma+\pi)=\pm e^{2\pi i\vartheta}\lambda(\tau, \sigma) \ . \eequ
The $\pm$ signs refer to Ramond and Neveu-Schwarz sectors respectively. These twisted periodicity 
conditions are analogous to those appearing in orbifolds, where the twisted states are 
closed strings whose centre of mass is stuck at the orbifold fixed point. 
In this case, it is the light-cone momentum that labels the twisted sectors whereas 
in the orbifold case there exists states of any transverse momentum in any twisted sector. 

The mode expansion for the non-trivial  modes is
\bequ Y(\sigma^+,\sigma^-)=i\sqrt{\frac{\alpha
'}{2}}\left\{\sum_{r \in \{{\mathbb{Z}}-\vartheta\}}\frac{\widetilde{\alpha}_r}{r}e^{-2ir\sigma^+} + \sum_{r\in\{{\mathbb{Z}}+\vartheta\}}\frac{\alpha_r}{r}e^{-2ir\sigma^-}\right\} \ , \label{yme} \eequ
\bequ
\lambda(\sigma^-)=\sqrt{2\alpha'}\left\{ \barr{c} \displaystyle{\sum_{r\in \left\{{\mathbb{Z}}+\vartheta \right\}}d_r e^{-2ir\sigma^-} \ ({\rm{Ramond}})} \spa{0.6}\\ \displaystyle{\sum_{r\in \left\{{\mathbb{Z}}+\vartheta +1/2\right\}}c_r e^{-2ir\sigma^-} \ ({\rm{Neveu-Schwarz}})} \earr \right.\ . \eequ
In the Ramond sector the modding is the same as for the superpartner. 
The mode expansion for the trivial modes is
\bequ 
\zeta^a(\sigma^+,\sigma^-)=\zeta^a_0+2\alpha' p^a\tau +i\sqrt{\frac{\alpha
'}{2}}\sum_{n \in \{{\mathbb{Z}}\backslash \{0\}\}}\left\{\frac{\widetilde{\zeta}_n^a}{n}e^{-2in\sigma^+} +\frac{\zeta_n^a}{n}e^{-2in\sigma^-}\right\} \ , \eequ
\bequ
\Psi^{\zeta^a}(\sigma^-)=\sqrt{2\alpha'}\left\{ \barr{c} \displaystyle{\sum_{n\in {\mathbb{Z}}}d_n^a e^{-2in\sigma^-} \ ({\rm{Ramond}})} \spa{0.6}\\ \displaystyle{\sum_{r\in \left\{{\mathbb{Z}}+1/2\right\}}c_r^a e^{-2ir\sigma^-} \ ({\rm{Neveu-Schwarz}})} \earr \right.\ . \eequ
Similar mode expansions hold for $\tilde{\psi}^X$ and $\tilde{\psi}^{\zeta^a}$ which are also free. 
The zero modes in the mode expansion for $v$ are 
\bequ
v=v_0+\alpha' q_-\sigma^-+\alpha'q_+\sigma^+ \ . 
\eequ
The $u,v$ coordinates are taken to be linear combinations of 
a timelike and a spacelike coordinate, where the 
spacelike coordinate is periodic with period $2\pi R$. Taking
\bequ u=\phi-t \ , \ \ \ \ v=\phi+t \  \eequ the
periodicity conditions for $t,\phi$ are 
\bequ
t(\tau,\sigma+\pi)=t(\tau,\sigma) \ , \ \ \ \ \ \phi(\tau,
\sigma+\pi)=\phi(\tau, \sigma)+ 2\pi Rw \ , 
\eequ 
where the
latter includes winding modes, with winding number $w\in {\mathbb{Z}}$. In terms of $u$ and $v$, the
periodicity conditions become 
\bequ u(\tau, \sigma+\pi)=u(\tau, \sigma)+
2\pi Rw \ , \ \ \ \ \ \ v(\tau, \sigma+\pi)=v(\tau, \sigma)+
2\pi Rw \ . \label{bc2} \eequ 

The worldsheet energy momentum tensor $T_{\pm \pm}$ has components
\bequ
\barr{c}
\displaystyle{T_{++}=\partial_+ u\partial_+ v+i\vartheta (X\partial_+\bar{X}-\bar{X}\partial_+X)+\partial_+X\partial_+\bar{X}+\partial_+\zeta^a\partial_+\bar{\zeta}^a} \spa{0.3}\\ \displaystyle{+\frac{i}{4}\left(\tilde{\psi}^X\partial_+\tilde{\psi}^{\bar{X}}+\tilde{\psi}^{\bar{X}}\partial_+\tilde{\psi}^{X}+\tilde{\psi}^{\zeta^a}\partial_+\tilde{\psi}^{\bar{\zeta^a}}+\tilde{\psi}^{\bar{\zeta^a}}\partial_+\tilde{\psi}^{\zeta}\right)} \ , \earr \label{tp} \eequ

\bequ
\barr{c}
\displaystyle{T_{--}=\partial_- u\partial_- v+ifp_-\alpha' (X\partial_-\bar{X}-\bar{X}\partial_-X)+\partial_-X\partial_-\bar{X}+\partial_-\zeta^a\partial_-\bar{\zeta}^a} \spa{0.3}\\ \displaystyle{\ \ \ \ \ \ \ \ \ \ \ \ \ \ +\frac{i}{4}\left(\psi^X\partial_-\psi^{\bar{X}}+\psi^{\bar{X}}\partial_-\psi^{X}+\psi^{\zeta^a}\partial_-\psi^{\bar{\zeta^a}}+\psi^{\bar{\zeta^a}}\partial_-\psi^{\zeta^a}\right)+\alpha'p_-f\psi^X\psi^{\bar{X}}} \ , \earr \label{tm} \eequ
and the classical Virasoro zero modes
\bequ
L_0=\frac{1}{4\pi\alpha'}\int_0^{\pi}d\sigma T_{--} \ , \ \ \ \ \widetilde{L}_0=\frac{1}{4\pi\alpha'}\int_0^{\pi}d\sigma T_{++} \ , \eequ
are given by
\bequ
L_0=L_0^B+\left\{\barr{l}L_0^R \\ L_0^{NS}\earr \right.  \ , \ \ \ \widetilde{L}_0=\widetilde{L}_0^B+\left\{\barr{l}\widetilde{L}_0^R \\ \widetilde{L}_0^{NS}\earr \right. \ , 
\eequ
where in the right moving sector we have
\bequ
L_0^B=\frac{\alpha'(p_-q_-+\vec{p}^2)}{4}+\frac{1}{2}\left\{\sum_{r\in\{{\mathbb{Z}}+\vartheta\}}\left(1-\frac{\alpha' fp_-}{r}\right)\alpha_r^*\alpha_r+\sum_{n\in {\mathbb{Z}}\backslash \{0\}}\zeta_n^{a*}\zeta_n^{a}\right\} \ ,
\eequ
\bequ
L_0^{R}=\frac{1}{2}\left\{\sum_{r\in\{{\mathbb{Z}}+\vartheta\}}(r-\alpha'fp_-)d_r^*d_r+\sum_{n\in{\mathbb{Z}}}rd_r^{a*}d_r^{a}\right\} \ , \eequ
\bequ
L_0^{NS}=\frac{1}{2}\left\{\sum_{r\in\{{\mathbb{Z}}+\vartheta+1/2\}}(r-\alpha' fp_-)c_r^*c_r+\sum_{r\in\{{\mathbb{Z}}+1/2\}}rc_r^{a*}c_r^{a}\right\} \ , \eequ
and in the left moving sector we have
\bequ
\widetilde{L}_0^B=\frac{\alpha'(p_+q_++\vec{p}^2)}{4}+\frac{1}{2}\left\{\sum_{r\in\{{\mathbb{Z}}-\vartheta\}}\left(1+\frac{\vartheta}{r}\right)\widetilde{\alpha}_r^*\widetilde{\alpha}_r+\sum_{n\in {\mathbb{Z}}\backslash \{0\}}\widetilde{\zeta}_n^{a*}\widetilde{\zeta}_n^{a}\right\} \ ,
\eequ
\bequ
\widetilde{L}_0^{R}=\frac{1}{2}\sum_{r\in{\mathbb{Z}}}r\{\widetilde{d}_r^*\widetilde{d}_r+\widetilde{d}_r^{*a}\widetilde{d}_r^a\} \ , \ \ \ \ \ \widetilde{L}_0^{NS}=\frac{1}{2}\sum_{r\in\{{\mathbb{Z}}+1/2\}}r\{\widetilde{c}_r^*\widetilde{c}_r+\widetilde{c}_r^{*a}\widetilde{c}_r^a\} \ . \eequ 
Equal time canonical (anti-)commutation relations for the target space coordinates 
\bequ
[P_X(\sigma), X(\sigma')]=-2i\delta(\sigma-\sigma') \ , \ \ \ \ \ \left\{P_{\psi}(\sigma),\psi^X(\sigma')\right\}=-2i\delta(\sigma-\sigma') \ , \eequ
are obeyed by choosing the following (anti-)commutation relations for the Fourier modes.
\bequ
[\widetilde{\alpha}_r,\widetilde{\alpha}_s^*]=2r\delta_{rs} \ , \ \ r,s\in \{{\mathbb{Z}-\vartheta}\} \ , \ \ \ \ \ [\alpha_r,\alpha_s^*]=2r\delta_{rs} \ , \  \ r,s\in \{{\mathbb{Z}+\vartheta}\} \ , \eequ 
\bequ
\{d_r,d_s^*\}=2\delta_{rs} \ , \ \ r,s\in \{{\mathbb{Z}+\vartheta}\} \ , \ \ \ \ \ \{c_r,c_s^*\}=2\delta_{rs} \ , \  \ r,s\in \{{\mathbb{Z}+\vartheta+1/2}\} \ . \eequ 
Similar relations hold for the fermions in the left moving sector and for the transverse coordinates modes. 
The quantum Virasoro zero modes are, using zeta function regularization\footnote{The $\zeta$-function regularization only works for the right moving sector.},
\bequ
L_0=\frac{\alpha'(p_-q_-+\vec{p}^2)}{4}+N+2fRw J \ , \ \ \ \widetilde{L}_0=\frac{\alpha'(p_+q_++\vec{p}^2)}{4}+\widetilde{N} \ , \eequ
where
\bequ
N=N^B+\left\{\barr{c}N^R \\ N^{NS}-1/2 \earr \right. \ ,\ \ \ \ \ \ \widetilde{N}=\widetilde{N}^B+\left\{\barr{c}\widetilde{N}^R \\ \widetilde{N}^{NS}-1/2 \earr \right. \ , \eequ
and the number operators are defined by
\bequ
N^B=\frac{1}{2}\left\{\sum_{r\in \{{\mathbb{Z}}^++\vartheta\}}\left(1-\frac{\vartheta}{r}\right)\alpha_{r}^*\alpha_{r}+\sum_{r\in \{{\mathbb{Z}}^+-\vartheta\}}\left(1+\frac{\vartheta}{r}\right)\alpha_{-r}\alpha_{-r}^*+\sum_{n\in {\mathbb{Z}}^+}[\zeta_{+n}^{a*}\zeta_{+n}^a+\zeta_{-n}^a\zeta_{-n}^{a*}]\right\} \ , 
\eequ
\bequ
N^R=\frac{1}{2}\left\{\sum_{r\in \{{\mathbb{Z}}^++\vartheta\}}\left(r-\vartheta\right)d_{r}^*d_{r}+\sum_{r\in \{{\mathbb{Z}}^+-\vartheta\}}\left(r+\vartheta\right)d_{-r}d_{-r}^*+\sum_{n\in {\mathbb{Z}}^+}n[d_{n}^{a*}d_{n}^a+d_{-n}^ad_{-n}^{a*}]\right\} \ , 
\eequ
\bequ
N^{NS}=\frac{1}{2}\left\{\sum_{r\in \{{\mathbb{Z}}^+_0+1/2+\vartheta\}}\left(r-\vartheta\right)c_{r}^*c_{r}+\sum_{r\in \{{\mathbb{Z}}^+-1/2-\vartheta\}}\left(r+\vartheta\right)c_{-r}c_{-r}^*+\sum_{r\in \{{\mathbb{Z}}^+_0+1/2\}}r[c_{r}^{a*}c_{r}^a+c_{-r}^ac_{-r}^{a*}]\right\} \ , 
\eequ
and the angular momentum operator is given by
\bequ
2J=\sum_{r\in\{{\mathbb{Z}}^+_0+\vartheta\}}\frac{\alpha_ {r}^*\alpha_ {r}}{r}-
\sum_{r\in\{{\mathbb{Z}}^+-\vartheta\}}\frac{\alpha_{-r}\alpha_{-r}^*}{r}
 +  \left\{\barr{c}\displaystyle{ \sum_{r\in\{{\mathbb{Z}}^+_0+\vartheta\}}
d_ {r}^*d_ {r}-
\sum_{r\in\{{\mathbb{Z}}^+-\vartheta\}}d_{-r}d_{-r}^*}\spa{0.3}\\
\displaystyle{\sum_{r\in\{{\mathbb{Z}}^+_0+1/2+\vartheta\}}
c_ {r}^*c_ {r}-
\sum_{r\in\{{\mathbb{Z}}^+-1/2-\vartheta\}}c_{-r}c_{-r}^*+1} \earr \right. \ .\eequ
The number operators for the left-movers are obtained from the above by introducing tilde modes and taking
$\vartheta\rightarrow -\vartheta$ in the bosonic part and $\vartheta=0$ in the fermionic sector. The number operators have integer eigenvalues for all states surviving GSO projection. The angular momentum operator has integer eigenvalues in the Ramond sector and half-integer in the Neveu-Schwarz sector. 

Introducing Kaluza-Klein momentum $m$ and energy $E$ we have
\bequ
\frac{m}{R}=\int_{0}^{\pi}d\sigma \frac{\partial {\mathcal{L}}}{\partial \dot{\phi}} = \frac{p_-+p_++q_-+q_+}{4}-2fJ \ , \eequ
\bequ
E=\int_{0}^{\pi}d\sigma \frac{\partial {\mathcal{L}}}{\partial \dot{t}} = \frac{p_-+p_+-q_--q_+}{4}+2fJ \ . \eequ
It follows that
\bequ
p_{\pm}=E+Q_ {\pm} \ , \ \ \ \ \ q_ {\pm}=Q_ {\pm}-E+4fJ \ .
\eequ
where the the right and left moving \lq momenta' are given by \bequ
Q_{\pm}=\frac{m}{R}\pm \frac{Rw}{\alpha'} \ . \eequ
The Virasoro zero modes can then be rewritten 
\bequ
L_0=\frac{\alpha'}{4}(-E^2+Q_-^2+\vec{p}^2)+N+\alpha'fJ(E+Q_+) \ , \eequ
\bequ
\widetilde{L}_0=\frac{\alpha'}{4}(-E^2+Q_+^2+\vec{p}^2)+\tilde{N}+\alpha'fJ(E+Q_+) \ . \eequ
The level matching condition yields
\bequ
N-\tilde{N}=mw \ , \eequ
and the mass shell condition gives the following dispersion relation.
\bequ
(E-2fJ)^2=\frac{4}{\alpha'}\widetilde{N}+(Q_++2fJ)^2+\vec{p}^2 \ . 
\eequ
Finally, we rewrite the Virasoro zero modes in a convenient form.
\bequ
L_0=\frac{\alpha'}{4}\left(-\Biggl(E-2fJ\Biggr)^2+\vec{p}^2+(Q_++2fJ)^2\right)-\frac{mw}{2}+N \ , \eequ
\bequ
\widetilde{L}_0=\frac{\alpha'}{4}\Biggl(-\left(E-2fJ\Biggr)^2+\vec{p}^2+(Q_++2fJ)^2\right)+\widetilde{N} \ . \eequ

\addtocounter{section}{1}
\setcounter{equation}{0}
\section*{Appendix D: $H_4$-algebra}

The G\"odel-type spacetime possesses an $H_4$ isometry group. The Heisenberg group is a generalization of the 
translation symmetry in flat spacetime. The $H_4$ currents are given by (see, for example, \cite{D'Appollonio:2003dr})
\begin{eqnarray}
K&=&-if\del_-u\ ,\quad 
J={1 \over f}\left(i\del_-v+2ifr^2\del_-\theta\right)\ ,\quad
P^{\pm}=\sqrt{2}e^{\mp 2ifu}\del_-\left(re^{\mp i\theta}\right)
\ ,\\
\widetilde{K}&=&if\del_+u\ ,\quad
\widetilde{J}=-{1 \over f}\left(i\del_+v-2ifr^2\del_+\theta
-4if^2r^2\del_+u\right)\ ,
\nn\\
&&\hspace{6cm}\widetilde{P}^{\pm}=\sqrt{2}e^{\pm 2ifu}\del_+
\left(re^{\mp i\theta}e^{\mp 2ifu}\right)\ .
\end{eqnarray}
In \lq light-cone' gauge, these currents become
\begin{eqnarray}
K&=&-i\alpha' fp_-\ ,\quad 
J={i \over f}\alpha' q_-+{i \over f}\del_-V_-
-(Y_-\del_-\bar{Y}_--\bar{Y}_-\del_-Y_-)\ ,\nn\\
&&P^{+}=\sqrt{2}e^{- 2i\alpha' fp_-\sigma^-}\del_-\bar{Y}_-
\ ,\quad
P^{-}=\sqrt{2}e^{2i\alpha' fp_-\sigma^-}\del_-Y_-\ ,\\
\widetilde{K}&=&i\alpha' fp_+\ ,\quad
\widetilde{J}=-{i \over f}\alpha' q_+-{i \over f}\del_+V_+
-(Y_+\del_+\bar{Y}_+-\bar{Y}_+\del_+Y_+)\ ,\nn\\
&&\widetilde{P}^{+}=\sqrt{2}e^{2i\alpha' fp_+\sigma^+}
\del_+\bar{Y}_+\ ,\quad
\widetilde{P}^{-}=\sqrt{2}e^{-2i\alpha' fp_+\sigma^+}
\del_+Y_+\ .
\end{eqnarray}
The Sugawara currents, $T=\half(P^+P^-+P^-P^+)+JK$ and $\widetilde{T}=\half(\widetilde{P}^+\widetilde{P}^-+\widetilde{P}^-\widetilde{P}^+)+\widetilde{J}\widetilde{K}$, indeed reproduce the stress-energy tensors up to normal ordering.

In the light-cone gauge, the full-fledged affine $H_4$ algebra is broken down to a smaller subalgebra,
\begin{eqnarray}
\left[P^+_m,P^-_n\right]&=&2\left(m-iK_0\right)\delta_{m+n,0}\ ,\quad
\left[J_0,P^{\pm}_m\right]=\mp iP^{\pm}_{m}\ ,\\
\left[\widetilde{P}^+_m,\widetilde{P}^-_n\right]&=&
2\left(m-i\widetilde{K}_0\right)\delta_{m+n,0}\ ,\quad
\left[\widetilde{J}_0,\widetilde{P}^{\pm}_m\right]
=\mp i\widetilde{P}^{\pm}_{m}\ .
\end{eqnarray}
Note that $P^{\pm}_m$ and $\widetilde{P}^{\pm}_m$ are simply the transverse oscillators, with the left-moving 
oscillators 
identified by $\widetilde{P}^{+}_{-m}=\widetilde{\alpha}^{\ast}_{m-\vartheta}$ 
and $\widetilde{P}^{-}_m=\widetilde{\alpha}_{m-\vartheta}$, while the right-moving 
oscillators are given by 
$P^+_{-(m+2fRw)}=\alpha^{\ast}_{m+\vartheta}$ and $P^-_{m+2fRw}=\alpha_{m+\vartheta}$. 
The presence of winding sectors changes the modding of some of the right moving modes relative to the
uncompactified case. Nonetheless, the subalgebra is satisfied.
The spectral flowed operators yield an isomorphic algebra.
\begin{equation}
P^{\pm '}_m=P^{\pm}_{m\mp a}\ ,\quad
K'_0=K_0-ia\ ,\quad
J'_0=J_0-ib\ ,\quad
L'_0=L_0-iaJ_0-ibK_0+ab\ ,
\end{equation}
The left moving algebra is invariant under an independent spectral flow, which takes the same form.
In the string quantization, the definition of the vacuum for $k-1<\vartheta_k<k$ was given by
\begin{equation}
\widetilde{\alpha}^{\ast}_{m-\vartheta_k}
\widetilde{\ket{\vartheta_k}}
=\widetilde{P}^+_{-m}\widetilde{\ket{\vartheta_k}}=0\ ,
\quad\mbox{for}\quad m<k\ ,\quad
\widetilde{\alpha}_{m-\vartheta_k}
\widetilde{\ket{\vartheta_k}}
=\widetilde{P}^-_m\widetilde{\ket{\vartheta_k}}=0\ ,
\quad\mbox{for}\quad m>k\ ,
\end{equation}
with a similar result for the right vacuum. 
The vacuum is not the highest weight state of the original current algebra. However, 
the vacuum is a highest weight state of the spectral flowed algebra: $\widetilde{P}^{+ '}_m=\widetilde{P}^{+}_{m-k}\ ,\widetilde{P}^{- '}_m=\widetilde{P}^{-}_{m+k}\ ,\widetilde{K}^{'}_0=\widetilde{K}_0-ik$. When quantizing the string, there are two points of
view one can take. One can either work with the same vacuum, and consider the spectral flowed Sugawara currents, or
one can work with the same Sugawara currents and use spectral flowed representations. In this paper, the latter viewpoint was
taken.

\end{document}